\documentclass[]{cit_lab_mfr}
\usepackage{hyperref}
\usepackage{cleveref}
\usepackage{verbatim}

\usepackage{wrapfig}  
\usepackage{graphicx}
\usepackage{floatrow}
\usepackage{subcaption}
\usepackage{listings}
\usepackage{algorithm}
\usepackage{wrapfig}
\usepackage{subfig}
\usepackage{epigraph}
\usepackage{subcaption} %

\usepackage[toc,page,header]{appendix}

\usepackage{minitoc}

\title{SAIL-Embedding Technical Report: Omni-modal Embedding Foundation Model}

\affiliation{%
\parbox{\textwidth}{\centering
\textbf{ByteDance Douyin SAIL Team, ~CUHK MMLab}
}}

\renewcommand{\thefootnote}{\fnsymbol{footnote}}
\renewcommand{\thefootnote}{\arabic{footnote}}  

\usepackage{xspace}
\newcommand{\modelname}{SAIL-Embedding\xspace}

\abstract{
Multimodal embedding models aim to yield informative unified representations that empower diverse cross-modal tasks. Despite promising developments in the evolution from CLIP-based dual-tower architectures to large vision-language models, prior works still face unavoidable challenges in real-world applications and business scenarios, such as the limited modality support, unstable training mechanisms, and industrial domain gaps. In this work, we introduce SAIL-Embedding, an omni-modal embedding foundation model that addresses these issues through tailored training strategies and architectural design. SAIL-Embedding supports multifaceted multimodal retrieval and classification by accommodating arbitrary modality inputs, including transcribed textual information, sampled visual semantics, and acquirable audio signals.
To enhance training robustness and scalability, we introduce the dynamic hard negative mining and adaptive multi-source data balancing to consolidate domain expertise and capture effective multimodal representations. In the optimization procedure, we propose a multi-stage training scheme to boost the multifaceted effectiveness of representation learning. Specifically, the content-aware progressive training aims to enhance the model's adaptability to diverse downstream tasks and master enriched cross-modal proficiency. The collaboration-aware recommendation enhancement training further adapts multimodal representations for recommendation scenarios by distilling knowledge from sequence-to-item and ID-to-item embeddings while mining user historical interests. Concurrently, we develop the stochastic specialization and dataset-driven pattern matching to strengthen model training flexibility and generalizability.
Experimental results demonstrate that SAIL-Embedding achieves state-of-the-art performance compared to other methods in item-to-item and query-to-item retrieval tasks across different intents. Furthermore, we provide comprehensive analysis and ablation studies to reveal the necessity of the proposed modules and components.
 In online experiments across various real-world scenarios integrated with our model, we observe a significant increase in Lifetime (LT), which is a crucial indicator for the recommendation experience. For instance, the model delivers the 7-day LT gain of +0.5\% in the Douyin-Selected scenario. Furthermore, through clustering quantification, the model is widely applied across diverse situations such as decentralization, recall, pre-ranking, and re-ranking. For the Douyin feed rank model, the match features produced by \modelname yield a +0.1\% AUC gain.
}
\date{\today}
\begin{document}
\maketitle

\begingroup 
\renewcommand{\thefootnote}{\fnsymbol{footnote}}
\endgroup

\begin{figure}
    \centering
    \includegraphics[width=1\linewidth]{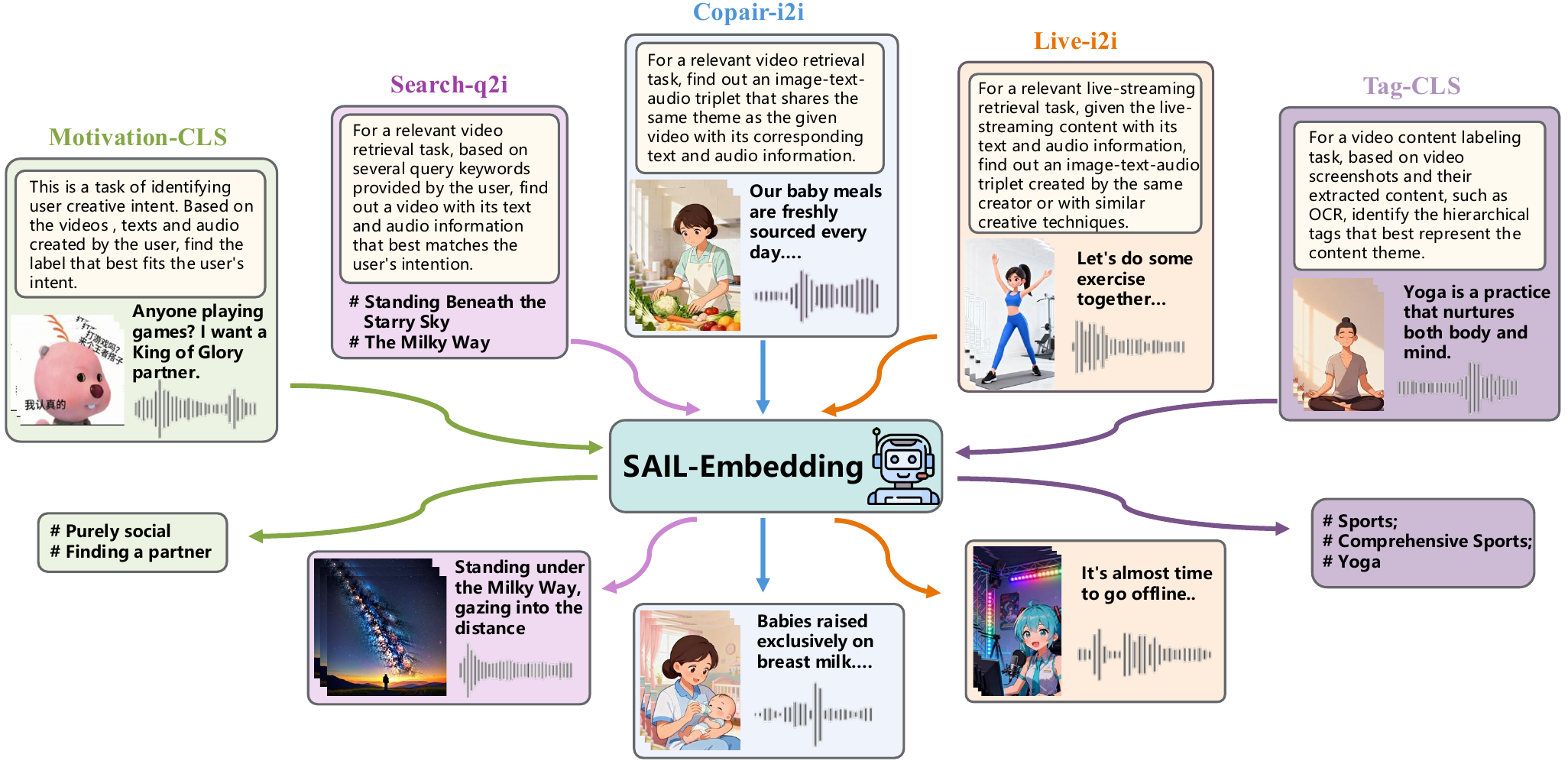}
    \caption{\textbf{SAIL-Embedding Capability Overview}. We present SAIL-Embedding, an omni-modal embedding model adapted from vision-language-audio models. SAIL-Embedding is capable of following instructions and performing various omni-modal embedding tasks, such as Motivation/Tag-CLS (Classification), Search-q2i (Query-to-Item Retrieval), and Copair/Live-i2i (Item-to-Item Retrieval).}
    \label{fig:demo}
\end{figure}

\section{Introduction}
\label{sec:intro}

Embedding models aim to generate meaningful dense vector representations of data. From the early exploration of distributed word representations such as Word2Vec~\cite{mikolov2013efficient} and GloVe~\cite{pennington2014glove} to the recent development of large language model (LLM)-based embedding models such as Gemini-Embedding~\cite{lee2025gemini} and Qwen-Embedding~\cite{zhang2025qwen3}, textual embedding models~\cite{Conan-embedding,Jina,Jina2} have become a fundamental component of natural language processing 
and have enabled a wide range of downstream applications, like retrieval-augmented generation (RAG)~\cite{izacard2020leveraging,jiang2023active} and information extraction~\cite{tang2025missing,mitra2017learning,karpukhin2020dense}.
With the rapid progress of multimodal learning~\cite{Guo2025Seed15VLTR, yang2022disentangled, yang2024MCIS, yang2022emotion, Wang2025RICOIA, Bai2023QwenVLAV,wang2024world,wang2025vgr,lei2025scalability}, multimodal embedding models~\cite{zhang2024gme,jiang2024vlm2vec,vlm2vec2,springer2024repetition,E5-v,Uniir,Vista,Mm-embed,MoCa} that map heterogeneous modalities into a unified vector space 
have also emerged as an active research direction, showing great promises in practical scenarios 
such as short-video,  image recommendation, as well as cross-modal search.

\begin{figure*}[t]
    \centering 
    \includegraphics[width=0.95\linewidth]{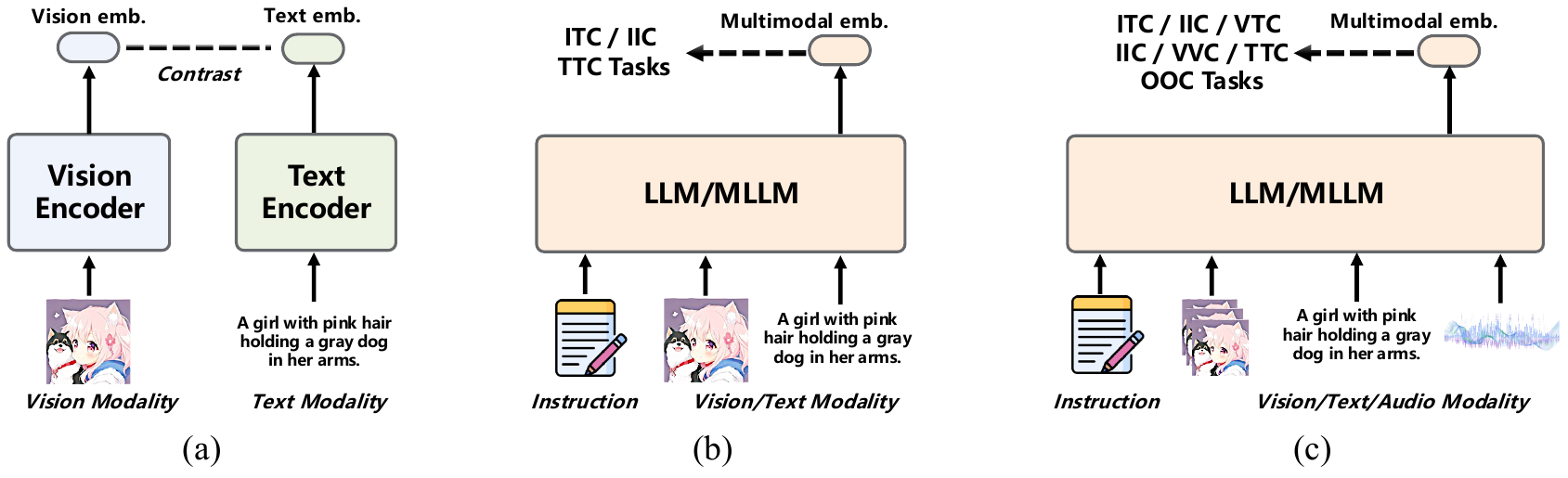}
    \caption{\textbf{Embedding Architecture Comparison}. (a) CLIP-like dual-tower embedding model architecture. (b) LLM/MLLM-based embedding model architecture. (c) SAIL-Embedding model architecture. In contrast, our model accommodates arbitrary modality inputs and can handle diverse downstream tasks. I, V, T, and O represent image, video, text, and omni-modality, respectively.
}
    \label{fig:intro}
\end{figure*}

Current multimodal embedding models generally follow two design paradigms, as illustrated in Figure~\ref{fig:intro}.  
The first paradigm, exemplified by CLIP~\cite{radford2021learning} and SigLIP~\cite{zhai2023sigmoid}, leverages paired multimodal data and employs two large encoders to process each modality independently, with either no fusion or only shallow fusion layers.  
This design is structurally simple and has proven highly effective for cross-modal retrieval.  
However, its reliance on shallow fusion restricts the expressiveness and semantic richness of the resulting embeddings.  
With the recent advances in large language models (LLMs) and multimodal LLMs (MLLMs), a second paradigm has gained increasing attention, as shown in Figure~\ref{fig:intro}(b).  
This line of work integrates LLMs/MLLMs to achieve deep semantic fusion across modalities~\cite{zhang2024gme,jiang2024vlm2vec,MoCa,Mm-embed}.  
Although still under active investigation, this paradigm has already exhibited strong representational capacity and is rapidly emerging as the de facto choice for a wide range of downstream tasks~\cite{gu2025breaking}.

However, when deployed for real-world scenarios, the models still face significant limitations that hinder their effectiveness.  
\textbf{(1) Limited Modalities:} Most existing methods rely on only two modalities—typically images and text—for unimodal or cross-modal retrieval. In contrast, industrial applications often require richer and more comprehensive multimodal representations. For example, a Douyin video contains diverse sources of information: visual signals from the cover frame and keyframes, textual cues from tags and captions, background music as audio, and spoken content transcribed via automatic speech recognition (ASR). Each modality contributes essential semantic information, and omitting or misinterpreting even one can severely impact downstream tasks such as recommendation and search, ultimately degrading user experience.  
\textbf{(2) Training Instability:} These models are typically built upon multimodal large language models, which require careful architectural design and optimization strategies to ensure stable and efficient training.  
Developing optimization strategies that unlock the practical values of incentive models within business is an indispensable key.
\textbf{(3) Industrial Domain Gap:} Many models are trained and evaluated primarily on open-source datasets, but they often underperform on domain-specific industrial data, such as expressive short videos on Douyin, where data distributions and task requirements differ substantially from academic benchmarks.

To this end, we propose \textbf{SAIL-Embedding}, a powerful omni-modal embedding model of the SAIL families~\cite{dong2025scalable,SAILViT}, along with tailored training strategies to ensure stable large-scale optimization and balance modality contributions. 
SAIL-Embedding can handle arbitrary combinations of inputs from vision, text, and audio modalities, yielding multi-dimensional representation vectors to fulfil diverse real-world business requirements.

To further enhance the model's robustness during the training process and improve its scalability to handle large-scale and diverse datasets, we introduce dynamic hard negative mining and adaptive multi-source data balancing. The dynamic hard negative mining helps the model focus on distinguishing challenging negative samples, thereby consolidating the model’s understanding of domain-specific knowledge and reducing the risk of misclassification caused by ambiguous samples. Meanwhile, the adaptive multi-source data balancing dynamically learns weights directly from the data distribution to reduce reliance on manual parameter tuning and maintain the trade-off between data quality and distributional diversity.

\modelname presents the multi-stage training procedure that learns unified representations across multiple dimensions through content- and collaboration-awareness. Specifically, content-aware progressive training gradually enhances the embeddings' discriminative power for diverse task demands and the generalization ability to handle unseen scenarios by leveraging diverse, semantically rich data resources. This process endows the model with comprehensive domain knowledge capabilities, including cross-modal semantic integration, scenario-specific content understanding, and complex concept relation.
During the collaboration-aware recommendation enhancement phase, we perform multi-dimensional interest-driven sequence-to-item distillation to incorporate the users' historical behavioural patterns into the multimodal representations. Subsequently, the combined ID-to-item distillation further aggregates user-specific preference signals within the recommendation system, thereby improving the accuracy of item recommendations.

Extensive experimental results are conducted on multiple benchmark datasets covering different application scenarios, demonstrating that the proposed SAIL-Embedding achieves state-of-the-art (SOTA) performance compared to other advanced baseline methods (\textit{i.e.}, traditional unimodal models, dual-tower fusion models, and large vision-language representation methods) in diverse downstream tasks. 

In online experiments, \modelname demonstrates its effectiveness in the recommender system of Douyin, which brings substantial gains through diverse application pathways. 
For cold-start scenarios, our model achieves a \textbf{0.05\%} LT gain in total by performing embedding-based item2item (i2i) recall along with engaging embeddings into the recommendation model. Specifically, \modelname improves the AUC of the cold-start model by \textbf{0.1\%} when deployed as a target side feature.
We also discretize embeddings info semantic tokens, which ultimately deliver \textbf{$\sim$0.03\%} LT gain across different stages of the recommender system, including recall, pre-ranking and re-ranking. Specifically, the AUC of the ranking model can be improved by \textbf{$\sim$0.1\%} with the engagement of both embeddings and sematic tokens.

\section{Related Works}

We have witnessed remarkable progress in multimodal learning in recent years~\cite{Bai2023QwenVLAV,  Guo2025Seed15VLTR, Wang2024LaDiCAD, Wang2024InstructAvatarTE, Bai2024UniEditAU, han2024towards,li2024toward, yang2024towards, chen2024efficiency, li2024unified}. 
Among various research directions, \emph{multimodal embedding learning} has emerged as a fundamental paradigm, aiming to project heterogeneous modalities---such as images, text, audio, and video---into a shared representation space. 
Such unified embeddings enable a broad range of downstream tasks, including cross-modal retrieval~\cite{Uniir,Vista}, video understanding~\cite{yang2024asynchronous,xu2025debiased}, and recommendation~\cite{Onerec}. 
Existing approaches can be broadly categorized into two families: 
(1) \emph{dual-tower architectures}, which employ independent modality-specific encoders to map each modality to the joint space; and 
(2) \emph{multimodal large language model (MLLM)-based architectures}, which integrate all modalities into a unified Transformer framework for joint modeling.

\subsection{Dual-Tower Multimodal Embedding Models}
Dual-tower architectures, pioneered by CLIP~\cite{radford2021learning}, employ separate encoders for each modality (\textit{e.g.,} a Vision Transformer~\cite{dosovitskiy2020image} for images and a Transformer-based language model~\cite{Vaswani2017AttentionIA} for text), projecting them into a common embedding space.
Training typically relies on large-scale contrastive learning, maximizing the similarity between paired samples and minimizing it for mismatched pairs.
This decoupled design allows each encoder to be precomputed and cached, enabling highly efficient retrieval at inference time and scaling well to billions of items.
Following CLIP, numerous extensions have been proposed.
ALIGN~\cite{Jia2021ScalingUV} scales the model capacity and dataset size to improve representation quality.  
AudioCLIP~\cite{Guzhov2021AudioclipEC} incorporates an additional audio branch, while CLIP4Clip~\cite{Luo2021CLIP4ClipAE} adapts the architecture for video-text retrieval by encoding temporal information. More recently, BLIP~\cite{Li2022BLIPBL} and BLIP-2~\cite{Li2023BLIP2BL} bridge dual-tower and fusion-style paradigms: BLIP adopts a bootstrapped pre-training strategy that unifies vision-language understanding and generation, whereas BLIP-2 introduces a lightweight Q-Former to better align visual features with frozen large language models.  
SigLIP~\cite{Zhai2023SigmoidLF} further improves contrastive training by replacing the softmax cross-entropy with a sigmoid loss, mitigating the inefficiencies of batch-dependent normalization and allowing more stable training on large-scale noisy datasets.  
Other works~\cite{Desai2020VirTexLV, Joshi2024DataEfficientCL} investigate robust pretraining on noisy web-scale data by leveraging advanced filtering, tokenization, and data curation strategies.
Despite their efficiency and scalability, dual-tower architectures generally fuse modalities only in the final embedding space, which limits their ability to capture fine-grained token-level interactions, temporal dynamics, or higher-level multimodal reasoning required in complex tasks.

\subsection{MLLM-based Multimodal Embedding Models}
In contrast, MLLM-based approaches~\cite{zhang2024gme,jiang2024vlm2vec,vlm2vec2,springer2024repetition,E5-v,Uniir,Vista,Mm-embed,MoCa} aim to integrate all modalities within a unified sequence modeling framework, leveraging the generative and reasoning capabilities of large language models. 
Typically, modality-specific encoders (e.g., visual or audio spectrogram encoders) transform raw inputs into embeddings, which are then aligned to the language token space via linear projections or learned adapters.
Representative works include VLM2Vec~\cite{jiang2024vlm2vec} and GME~\cite{zhang2024gme}. 
VLM2Vec~\cite{jiang2024vlm2vec} generates fixed-dimensional embeddings for arbitrary combinations of images and text under task instructions, building upon Phi-3.5-V~\cite{Abdin2024Phi3TR}. 
Its successor, VLM2Vec-v2~\cite{vlm2vec2}, further extends the framework to support videos and visual documents. 
GME~\cite{zhang2024gme} constructs an MLLM-based dense retriever to enable unified cross-modal search. 
Subsequent research has sought to improve this paradigm: 
mmE5~\cite{Chen2025mmE5IM} leverages synthetic datasets for stronger multilingual performance, 
MoCa~\cite{MoCa} introduces bidirectional attention through continual pre-training to enhance scalability with both model size and training data, 
UniMoCo~\cite{Qin2025UniMoCoUM} proposes a modality-completion module that infers visual features from text to address modality-missing issues, 
UniME~\cite{Gu2025BreakingTM} employs discriminative knowledge distillation from a powerful LLM teacher to improve the embedding quality of the language component,
and NoteLLM-2~\cite{Zhang2024NoteLLM2ML} explores leveraging multimodal large representation models for recommendation.

By naturally modeling cross-modal dependencies through Transformer-based self-attention~\cite{Vaswani2017AttentionIA}, this paradigm enables deeper semantic understanding, contextual reasoning, and compositionality—capabilities that dual-tower models often lack. 
Nonetheless, most existing methods remain constrained to image and text modalities, falling short of fully supporting omni-modal understanding.

\section{Methodology}

In this section, we present the construction of the proposed \modelname model. 
We first describe the collected datasets in \textsection~\ref{sec: data}, together with strategies for balancing heterogeneous data and a hard negative mining method to strengthen representation learning. 
We then introduce the model architecture in \textsection~\ref{sec: arch}, highlighting the fusion of different modalities and the use of prompts to fully leverage the capabilities of multimodal large language models. 
Finally, we define the training objectives in \textsection~\ref{sec:method_train} and outline several techniques designed to improve the training effectiveness.

\subsection{Data Preparation and Preprocessing} \label{sec: data}
\subsubsection{Recommendation-aware Data Construction}
The data construction goal for SAIL-Embedding is to enable the model to provide omni-modal understanding capabilities for practical applications, meeting recommendation demands across diverse scenarios,  such as Douyin videos and Douyin live.
We curate a large-scale dataset of over 10B samples, with statistical details provided in Table~\ref{tab_data}.
Mostly, each sample is a pair consisting of a query and a target for CLIP-like contrastive learning. The query can be a video or just a couple of words, as well as the target. 
Different training datasets are designed with specific philosophies to encompass diverse content and collaboration semantics.
To this end, the meta-task categories for training data are summarized as follows:

\begin{enumerate}
    \item \textbf{Item-to-Item Retrieval}: Typically, an item serves as a video containing omni-modal information, encompassing vision, text, and audio modalities. This type of task performs retrieval of candidate items based on various requirements, including user behaviors, video summaries, semantic IDs, and specific business applications.
    \item \textbf{Query-to-Item Retrieval}: In this task, the query only contains short texts for item search. And the target is a video that is mostly clicked by users from the search results list. Based on these pairs, we further use rule-based and LLM-based methods to generate similarity scores between each query and target, labeling for tasks like the COSENT training~\cite{kexuefm-8847}. 
    \item \textbf{Classification}: The query is typically a given item, with the target being the corresponding multi-level tags.  Inspired by UniCL~\cite{yang2022unified} and iCLIP~\cite{wei2023iclip}, we transform classification datasets into item–label text pairs. This type of task spans a wide range of scenarios to serve different tag recognition needs, such as user motivation and typical image-text classification.
\end{enumerate}

\begin{table}[t]
\centering
\caption{\textbf{SAIL-Embedding Training Data Statistics.} V, T, and A represent vision, text, and audio modalities, respectively. The training data encompasses multi-faceted retrieval tasks across queries and items with omni-modal information, as well as multi-level label classification tasks. ``RSDF'' means the Recommendation-side Dense Features.}
\resizebox{\linewidth}{!}{%
\begin{tabular}{cccccc}
\toprule[1.5pt]
\textbf{Meta Task }                                                                          & \textbf{Data Partition}  & \textbf{Design Philosophy}                                                                                                                                                      & \textbf{Query} & \textbf{Target} & \textbf{Data Magnitude} \\ \midrule
\multirow{7}{*}{\begin{tabular}[c]{@{}c@{}}Item to Item\\  Retrieval\end{tabular}}  & Copair-i2i      & User consumption behaviour-based pair                                                                                                                                  & V+T+A & V+T+A  & 2.9B \\
                                                                                    & Search-i2i      & User search behaviour-based pair                                                                                                                                       & V+T+A & V+T+A  & 0.6B \\
                                                                                    & Live-i2i        & Live streaming-based behavioral copair                                                                                                                                        & V+T+A & V+T+A  & 1.4B \\
                                                                                    & Summary-i2i   & Ngram-based filtering pair                                                                                                                                             & V+T+A & V+T+A  & 1.1B \\          
                                                                                    & Hashtag-i2i         & User submission-based pair                                                                                                                                      & V+T+A & V+T+A  & 0.05B \\          
                                                                                    & RSDF-i2i        & RSDF similarity-based filtering pair                                                                                                                                     & V+T+A & V+T+A  & 0.3B \\          
                                                                                    & ID-i2i          & ID clustering-based pair                                                                                                                                               & V+T+A & V+T+A  & 0.08B \\ \midrule
\multirow{3}{*}{\begin{tabular}[c]{@{}c@{}}Query to Item \\ Retrieval \\ \vspace{0.5pt}\end{tabular}} & Search-q2i      & Search query and click-based pair                                                                                                                                      & T     & V+T+A  & 1.8B \\        
                                                                                    & Score-q2i       & \begin{tabular}[c]{@{}c@{}}Pairs of queries and targets with labeled similarity scores for downstream training\end{tabular} & T     & V+T+A  & 0.6B \\ \midrule 
Classification                                                                      & CLS             & Item information and corresponding multi-level tags                                                                                                                    & V+T+A & T      & 3.1B \\ \bottomrule[1.5pt]
\end{tabular}
}
\label{tab_data}
\end{table}

\subsubsection{Dynamic Hard Negative Mining}
In contrastive learning, the effectiveness of representation learning heavily depends on the quality of both positive and negative samples. While random negatives are often abundant, they tend to be semantically dissimilar to the query, making the discrimination task trivial and limiting the model's ability to capture fine-grained distinctions. Hard negatives---samples that are challenging to distinguish from positives due to high semantic similarity---play a crucial role in improving model robustness and retrieval performance. However, the notion of ``hard'' is dataset- and task-dependent, and applying a fixed global similarity threshold often leads to suboptimal results. To address this, we propose a dynamic hard negative mining strategy that adaptively determines the optimal similarity threshold for each dataset.

Formally, let $\mathcal{P} = \{(q_i, t_i)\}_{i=1}^{N}$ denote the set of positive pairs, where $q_i$ is a query and $t_i$ is its corresponding target. Negative pairs are constructed via a Cartesian product excluding positives:  
\[
\mathcal{N} = \{(q_i, t_j) \mid i \neq j, (q_i, t_j) \notin \mathcal{P}\}.
\]

We then compute cosine similarity scores $s_{ij} = \cos(q_i, t_j)$ for all negative pairs $(q_i, t_j) \in \mathcal{N}$. Merging positive and negative pairs, we obtain a dataset of $(s, y)$, where $y \in \{1 \text{ (positive)}, 0 \text{ (negative)}\}$.
For each candidate threshold $\lambda$, the binary prediction is defined as:
\[
\hat{y}_{ij}(\lambda) =
\begin{cases}
1, & s_{ij} \ge \lambda, \\
0, & s_{ij} < \lambda.
\end{cases}
\]

Precision and recall are computed accordingly, and the F1 score is defined as:
\[
\text{F1}(\lambda) = \frac{2 \cdot \text{Precision}(\lambda) \cdot \text{Recall}(\lambda)}{\text{Precision}(\lambda) + \text{Recall}(\lambda)}.
\]

The optimal similarity threshold is selected by:
\[
\lambda^* = \arg\max_\lambda \text{F1}(\lambda),
\]

which is then used to identify and mine hard negatives dynamically during training. This adaptive strategy ensures dataset- and task-specific selection of challenging negatives, improving contrastive representation learning and downstream retrieval performance.

Once $\lambda^*$ is determined, we construct the hard negative set $\mathcal{H}$ by selecting samples whose similarity scores are less than $\lambda^*$ yet among the highest below this threshold:
\begin{equation}
\mathcal{H} = \left\{ x_i \mid s(x_i) < \lambda^*,\  s(x_i) \ \text{is among the highest below} \ \lambda^* \right\},
\end{equation}
where $s(x_i)$ denotes the similarity score of sample $x_i$. 
This filtering step removes overly similar samples, which we regard as false negatives.
For each query $q_i$, we sample positives from $\mathcal{P}$, hard negatives from $\mathcal{H}$ and in-batch random negatives from $\mathcal{N}$. The contrastive loss is then defined as:
\[
\mathcal{L}_{\text{contrast}} = - \sum_{i=1}^N \left[ \log \frac{\exp(s_{ii}/\tau)}{\exp(s_{ii}/\tau) + \sum_{(q_i,t_j) \in \mathcal{H}} \exp(s_{ij}/\tau) + \sum_{(q_i,t_j) \in \mathcal{N}} \exp(s_{ij}/\tau)} \right],
\]
where $s_{ii}$ is the similarity of the positive pair, $\tau$ is a temperature hyper-parameter. This formulation adaptively enforces discrimination against semantically close but incorrect targets, thereby improving robustness and generalization of the learned representations.

\begin{figure}
    \centering
    \includegraphics[width=1\linewidth]{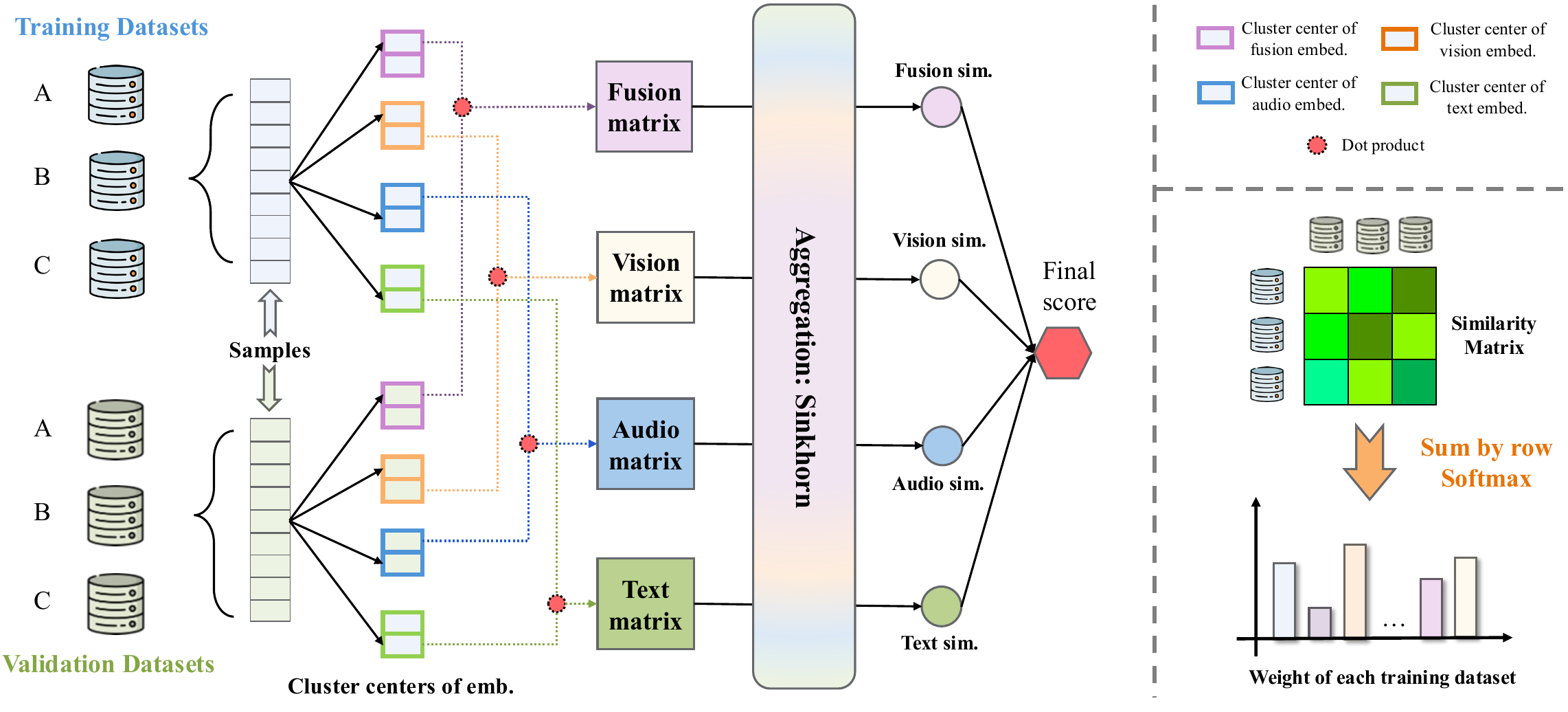}
    \caption{\textbf{Illustration of Adaptive Multi-Source Data Balancing.} For both training and validation datasets, we compute embeddings for different modalities and cluster them. We then construct a similarity matrix to represent the distance between training and validation clusters, which is further aggregated into a single similarity score. Based on these similarity values, each training dataset is assigned a weight using a Softmax function.
}
    \label{fig:placeholder}
\end{figure}

\subsubsection{Adaptive Multi-Source Data Balancing}
\label{amsdb}

Conventional multi-source training pipelines often rely on manually assigned dataset mixing ratios, determined by subjective expertise and task intuition. Such heuristic configurations are difficult to validate empirically and may lead to suboptimal generalization. We introduce an adaptive weighting framework that learns dataset-specific sampling weights directly from the data distribution, rather than from human-designed heuristics. The core idea is to measure the semantic similarity between high-quality benchmark validation sets and pre-training datasets, and to translate this similarity into flexible sampling weights for multi-source training. Compared with hard filtering approaches that remove entire samples based on a single-instance similarity score, our method performs \emph{soft selection} at the dataset level, preserving distributional diversity while maintaining overall data quality. This prevents overfitting to the benchmark domain and avoids distribution collapse, thereby improving generalization to unseen tasks.

Our pipeline works as illustrated in Figure~\ref{fig:placeholder}. We first construct validation subsets from the training datasets with distributions similar to the downstream test tasks. 
Given multiple training datasets and one or more benchmark validation datasets, we extract embeddings using an early version of our model. 
To reduce computation, we randomly sample approximately 10k samples per dataset and apply $k$-means clustering to obtain cluster centroids. 
For each dataset pair, we compute a cosine similarity matrix $C$ between their centroids. 
We then reduce $C$ to a scalar similarity score via the Sinkhorn algorithm~\cite{cuturi2013sinkhorn}, \textit{i.e.}, a weighted sum $\sum P \odot C$, where $P$ is a transport matrix derived from the distance matrix $1 - C$ using the Sinkhorn algorithm. 
This formulation assigns higher weights to more similar clusters. 
Next, we apply a ``fusion first'' strategy to determine the optimal modality combination, preferring fused modalities when available, and otherwise selecting unimodal representations. 
Finally, we compute the similarity between each training set $M_i$ and each benchmark $N_j$, producing an $m \times n$ similarity matrix. 
Averaging and normalizing across benchmarks yields the final adaptive sampling weight for each training set.

This method offers three benefits: (1) it learns weights directly from data distributions, reducing reliance on subjective manual tuning; (2) it is modular and can be generalized to other multi-source learning settings; (3) it maintains a balance between high data quality and distributional diversity, mitigating overfitting and enhancing robustness. 

\begin{figure*}[t]
    \centering   
    \includegraphics[width=1\linewidth]{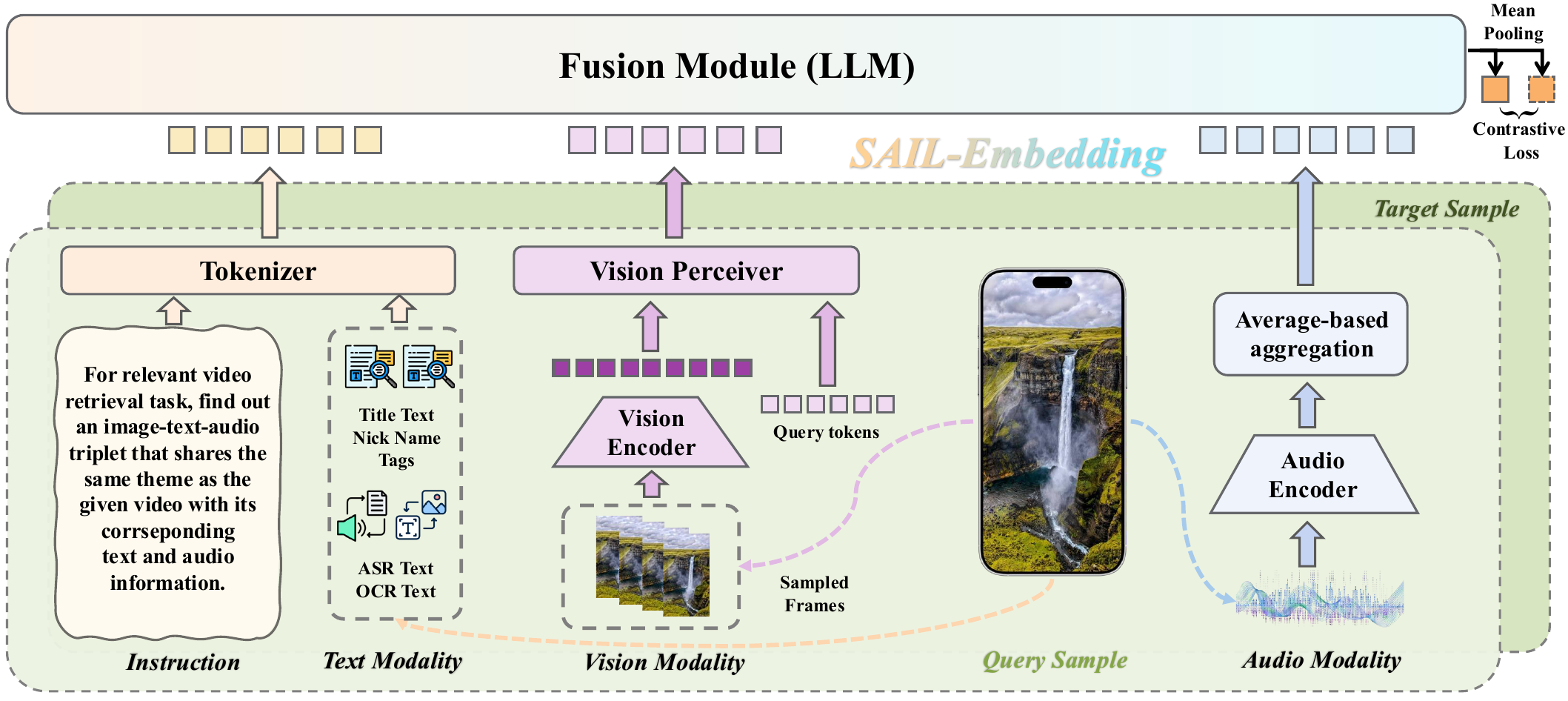}
    \caption{\textbf{Overview of Our \modelname Model.} For each sample, we extract relevant textual information (e.g., titles, OCR text) and combine it with instructions to obtain textual tokens. Visual and audio information are encoded using dedicated encoders, and the resulting multimodal tokens are concatenated and passed through a fusion module to obtain a unified multimodal embedding. The system is trained by contrasting embeddings from query and target samples.
}
    \label{fig:method}
\end{figure*}

\subsection{Architecture Design} \label{sec: arch}

\subsubsection{Overall Architecture}

The core idea of \modelname is to transform heterogeneous multimodal information into a unified embedding space, enabling robust cross-modal understanding and retrieval. As illustrated in Figure~\ref{fig:method}, given an input sample $x$ containing audio $a$, visual $v$, and textual $t$ signals, our framework leverages a large language model (LLM)~\cite{yang2025qwen3,team2024qwen2} as the central reasoning and integration backbone, which is warmed up following the work~\cite{dong2025scalable}.

For the text modality, we adopt conventional preprocessing pipelines, including tokenization and mapping each token to its corresponding word embedding via a trainable embedding layer.  
For visual and audio modalities, we follow the ``foreign language'' metaphor: each non-text modality is processed by a modality-specific encoder, namely $\mathcal{E}_{v}$ for vision and $\mathcal{E}_{a}$ for audio, to project their raw features into a natural language-compatible embedding space.  
These embeddings are then aligned in both dimension and semantics before being fed into the LLM for multimodal fusion.  
The final representation is extracted from the output of the LLM using mean pooling over all token embeddings.

This design allows for flexible integration of diverse modalities and provides a unified interface for knowledge transfer from large pre-trained language models to multimodal scenarios, without requiring extensive architecture modifications.

\subsubsection{Text Tokenizer}
\begin{figure}[t]
    \centering
    \includegraphics[width=1\linewidth]{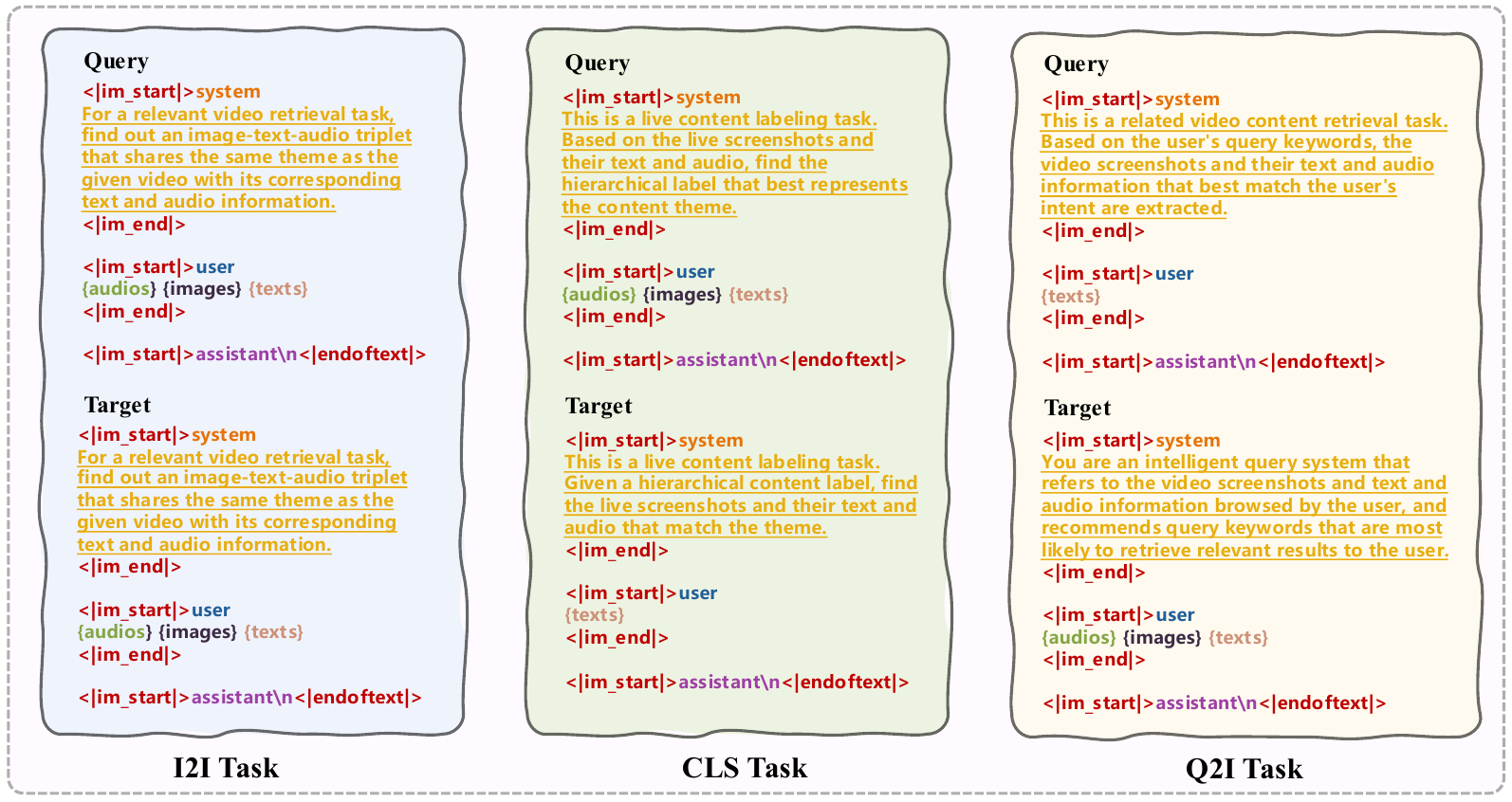}
    \caption{\textbf{Illustration of instructions for different tasks.} We design task-specific instructions by explicitly defining the task and its objective. Audio, image, and text tokens are then provided jointly to obtain the final results. For query and target tokens, modality-specific adaptations are applied to accommodate their respective modality combinations.
}
    \label{fig:instruct}
\end{figure}

Real-world short videos contain abundant textual information, such as titles, tags, author labels, OCR texts, and ASR texts. The mainly used fields are described as follows:
\begin{enumerate}
    \item \textbf{Title}: Title text information of the short video.
    \item \textbf{OCR Text}: Textual information recognized from the short video frames.
    \item \textbf{ASR Text}: Textual information obtained by converting audio from the short video.
    \item \textbf{Nickname}: The author's nickname associated with the short video.
    \item \textbf{Tags}: Text labels generated by tagging models for each item.
\end{enumerate}
We organize these signals into a unified textual format and concatenate them for downstream modeling.
Specifically, the title, ASR, and OCR text from the query and candidates, along with additional nickname and tags, are selected as the text-modal data. The processing workflow first performs cleaning and deduplication, followed by the random dropping of partial fields in practice to accommodate scenarios where original fields are missing during online deployment.

Moreover, to effectively leverage the knowledge of the multimodal language model, we design prompts that guide the model to process multimodal information more effectively. As illustrated in Figure~\ref{fig:instruct}. Each prompt consists of three parts. The first is the system prompt, which defines the task and specifies its objective, helping the model better interpret and represent subsequent content. Next, we insert the extracted image, text, and audio tokens as user-provided information. Finally, we append the \texttt{assistant} symbol to indicate that the model should generate an answer. In addition, since tasks vary, the instructions for query and target samples are carefully designed to differ, enabling the system to interpret the provided information appropriately for each case.

\subsubsection{Vision Encoding Module}

For the visual encoding, we adopt a Vision Transformer (ViT)-based backbone~\cite{SAILViT}. 
In the case of video data, each frame is independently patchified, generating a sequence of spatio-temporal tokens. 
We use a patch size of 14 and resize all frames to a uniform resolution to address resolution variation. 
Although such dense tokenization preserves fine-grained visual details, it also produces an excessive number of tokens, especially for high-resolution or long-duration videos, thereby imposing a substantial computational burden on downstream fusion.

To address this challenge, we introduce a \textit{Visual Perceiver} module, inspired by perceiver architectures~\cite{Jaegle2021PerceiverGP}, serving as a learnable bottleneck for token reduction. Concretely, we concatenate the visual tokens with $N_q = 16$ learnable latent query tokens, feed them into a Transformer block, and only retain the query token embeddings as the condensed visual representation. This mechanism preserves essential semantic content while significantly reducing sequence length, thereby improving both efficiency and scalability.

\subsubsection{Audio Encoding Module}
The audio modality processing typically has effective candidates, such as Qwen-audio~\cite{chu2023qwen}, Whisper~\cite{Whisper}, and CLAP models~\cite{elizalde2023clap}.
We empirically employ the CLAP model, which is faster than models like Whisper, to balance the efficiency of online deployment with mitigating the long-tail distribution issue in post-sampling audio sequence lengths. This method can extract high-level acoustic semantics to obtain discriminative audio tokens.
The pipeline is designed to accommodate diverse audio clip lengths:
\begin{enumerate}
    \item \textbf{Short audio} ($\leq 10$s): We apply a \emph{repeat-and-pad} operation to normalize the length before feature extraction, producing a single $1 \times \text{dim}$ audio token.
    \item \textbf{Long audio} ($> 10$s): We segment the waveform into consecutive non-overlapping 10s chunks, extract features for each chunk with CLAP, and aggregate them (via mean pooling) into a unified $1 \times \text{dim}$ representation.
\end{enumerate}
The resulting single audio token is appended to the multimodal token sequence before fusion, ensuring a consistent representation format regardless of the clip length.

\subsubsection{Fusion Module}

Let $\mathbf{T}_{a}$, $\mathbf{T}_{v}$, and $\mathbf{T}_{t}$ denote the token sequences obtained from the audio, visual, and textual modalities, respectively.  
We concatenate the unimodal token sequences to form a single multimodal sequence  
$\mathbf{T} = [\mathbf{T}_{a}; \mathbf{T}_{v}; \mathbf{T}_{t}]$,  
where $[\cdot ~; \cdot]$ denotes concatenation along the token dimension.  The fused sequence $\mathbf{T}$ is then fed into a multimodal Transformer fusion module  
(\textit{i.e.}, an LLM backbone~\cite{yang2025qwen3}), denoted as $\mathcal{F}(\cdot)$,  
which performs cross-modal reasoning via self-attention layers.  
To strengthen information exchange across modalities and capture long-range dependencies,  
we employ a bi-directional attention mechanism.  
The resulting hidden states are further normalized by a \textit{tanh} activation,  
ensuring embeddings remain within a bounded range, and subsequently aggregated by mean pooling  
to form the final multimodal embedding:  
\[
\mathbf{z} = \text{Meanpool}\big(\tanh(\mathcal{F}(\mathbf{T}))\big).
\]

\subsection{Training Strategies} \label{sec:method_train}

\subsubsection{Content-Aware Progressive Training}
\begin{figure}
    \centering
    \includegraphics[width=1\linewidth]{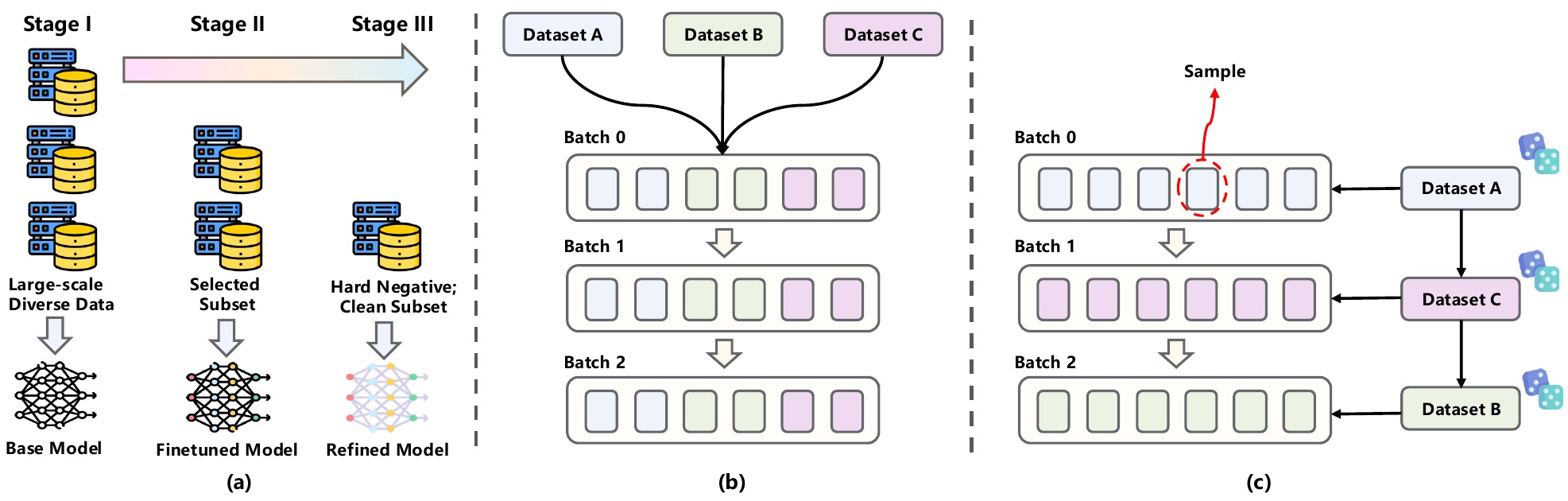}
    \caption{\textbf{Illustration of Training Techniques}. (a) Our progressive training framework gradually shifts from larger, diverse datasets to smaller, domain-specific datasets, balancing general-world knowledge with downstream specialization while maintaining training stability. (b) The conventional training method mixes heterogeneous data into a single batch. (c) Our stochastic specialization training randomly selects a dataset at each iteration to enhance robustness and specialization.
  }
    \label{fig:training}
\end{figure}

To adapt general-purpose large models to specific downstream tasks, we adopt a content-aware progressive training framework. As illustrated in Figure~\ref{fig:training}, each stage uses datasets with different scales and characteristics: earlier stages employ larger and more diverse datasets, while later stages focus on higher-quality data that closely matches the downstream task requirements. Specifically, in the first stage, we train a base model on large-scale and diverse datasets to acquire fundamental multimodal representation capabilities. Next, we fine-tune the model on a subset of datasets more aligned with the target downstream tasks. To further enhance the model's ability to capture fine-grained distinctions, we construct hard negatives and perform an additional fine-tuning stage to obtain the refined model. This progressive training framework enables the model to balance general-world knowledge with downstream domain-specific knowledge, while maintaining training stability.

\subsubsection{Loss Definition}

Our training objective follows the \emph{contrastive learning} paradigm to jointly optimize the multimodal embedding space. The overall loss integrates four complementary components: (i) a Noise-Contrastive Estimation (NCE) loss for global alignment, (ii) a COSENT loss~\cite{kexuefm-8847} for fine-grained ranking, (iii) a multimodal In-Context Learning (mICL) loss~\cite{Zhang2024NoteLLM2ML} to enhance modality-specific discrimination, and (iv) a late fusion loss~\cite{Zhang2024NoteLLM2ML} to balance visual and textual contributions.  

Given a query embedding $e_q$ and its positive target embedding $e_t^+$, the NCE loss enforces global alignment by contrasting against in-batch negatives:
\begin{equation}
\mathcal{L}_{\mathrm{nce}} = - \log \frac{\exp \left( \cos(e_q, e_t^+) / \tau \right)}
{\exp \left( \cos(e_q, e_t^+) / \tau \right) + \sum_{i=1}^B \exp \left( \cos(e_q, e_t^-) / \tau \right)},
\end{equation}
where $\tau$ is a learnable temperature parameter, initialized differently for each task and each dataset to account for distributional variations.

In real-world applications, storage and computational budgets impose strict constraints on the dimensionality of embedding vectors, making the deployment of high-dimensional representations prohibitive due to memory cost and retrieval latency. To overcome this limitation, we employ \emph{Matryoshka Representation Learning} (MRL)~\cite{Kusupati2022MatryoshkaRL}, which enforces multi-granularity supervision on embeddings of different sizes. 
Specifically, the full $1536$-dimensional embedding is sliced into multiple contiguous sub-vectors of sizes $768$ and $128$, and each slice, together with the full embedding, is optimized using the same InfoNCE loss objective:
\[
\mathcal{L}_{\mathrm{nce-mrl}} = \mathcal{L}_{\mathrm{nce}}(\mathbf{z}_{128}) 
+ \mathcal{L}_{\mathrm{nce}}(\mathbf{z}_{768}) 
+ \mathcal{L}_{\mathrm{nce}}(\mathbf{z}_{1536}),
\]
where $\mathbf{z}_{d}$ denotes the sub-embedding of dimensionality $d$. 
This ``nested'' training paradigm ensures that all sub-embeddings preserve strong discriminative power, enabling the system to dynamically trade off accuracy and efficiency at inference time without retraining. Such flexibility is particularly critical for deploying resource-constrained multi-modal embedding models.

For text-oriented pre-training datasets, we incorporate a cosine-similarity–based ranking loss (COSENT)~\cite{kexuefm-8847} to enforce fine-grained orderings:
\begin{equation}
\mathcal{L}_{\mathrm{cosent}} = \log \Bigg( 1 + \sum_{\mathrm{sim}(i,j) > \mathrm{sim}(k,l)} 
\exp \Big( \tfrac{\cos(e_k, e_l) - \cos(e_i, e_j)}{\tau} \Big) \Bigg).
\end{equation}

To mitigate modality imbalance, we adopt the multimodal In-Context Learning (mICL)~\cite{Zhang2024NoteLLM2ML}. Instead of compressing multimodal inputs into a single embedding, we separate them into modality-specific embeddings (visual $\mathbf{n}_v$ and textual $\mathbf{n}_t$), and apply in-batch contrastive learning to encourage consistent alignment across modalities:
\begin{equation}
\mathcal{L}_{\mathrm{micl}} = \mathcal{L}(\mathbf{n}_v, \mathbf{n}_v^+) + \mathcal{L}(\mathbf{n}_t, \mathbf{n}_t^+),
\end{equation}
where each term follows the NCE form but within its respective modality.  

In addition, we integrate a late fusion mechanism~\cite{Zhang2024NoteLLM2ML} to preserve visual fidelity. Given visual embedding $\mathbf{v}$ and multimodal embedding $\mathbf{n}_m$, a gated fusion module learns to adaptively combine them:
\begin{equation}
\mathbf{z} = \sigma \big(W[\mathbf{v}, \mathbf{n}_m] + b\big), \quad
\hat{\mathbf{n}}_m = \mathbf{z} \odot \mathbf{v} + (1-\mathbf{z}) \odot \mathbf{n}_m,
\end{equation}
where $\sigma$ denotes the sigmoid function. We then introduce a late-fusion contrastive loss $\mathcal{L}_{\mathrm{lf}}$ over $\hat{\mathbf{n}}_m$ to reinforce multimodal consistency.  
The final training loss is a weighted combination:
\begin{equation}
\mathcal{L} = \mathcal{L}_{\mathrm{nce-mrl}} + \lambda \, \mathcal{L}_{\mathrm{cosent}} + \alpha \, \mathcal{L}_{\mathrm{micl}} + \beta \, \mathcal{L}_{\mathrm{lf}},
\end{equation}
where $\lambda$, $\alpha$, and $\beta$ are balancing hyperparameters.  

\subsubsection{Stochastic Specialization Training}
In multi-domain training scenarios, as illustrated by Figure~\ref{fig:training} (b), data from heterogeneous datasets are often mixed within each iteration to ensure balanced utilization. However, such domain mixing inevitably fragments the effective batch size for each individual dataset, leading to small per-domain batches and elevated gradient variance. Moreover, synchronizing features or statistics across different domains within a single iteration introduces additional communication and processing overhead, which becomes more severe as the number of datasets grows.

To address this, we propose \emph{Stochastic Specialization Training}, a strategy inspired by meta-learning schemes that improves both supervision focus and computational efficiency. Instead of sampling from all datasets in every iteration, our method stochastically selects a single dataset according to a predefined probability distribution based on methods introduced in Section \ref{amsdb}, and draws the entire batch from it. Across training, all datasets are still visited, but each iteration specializes on one domain, yielding larger per-domain batch sizes while keeping the global batch size unchanged.

This specialization reduces gradient variance within each training step, simplifies the iteration logic, and eliminates the need for dataset-specific processing or inter-domain communication inside the iteration. Furthermore, the approach exhibits strong scalability: adding a new dataset requires only its dataset-specific configuration, without modifying the overall batching or communication pattern.

\subsubsection{Dataset-Driven Pattern Matching}
To address the challenge of heterogeneous modality availability and imbalance across datasets, we design a \emph{modality-aware matching strategy} that unifies various contrastive objectives under a general query-to-target framework. Specifically, we generalize the CLIP objective beyond the canonical image-to-text setting to a comprehensive \emph{modality-to-modality} paradigm, where any modality can serve as the query while the others become potential targets. This formulation enables flexible alignment tasks such as Image-to-Text Contrastive (ITC), Image-to-Image Contrastive (IIC), Video-to-Text Contrastive (VTC), Video-to-Video Contrastive (VVC), Text-to-Text Contrastive (TTC), and Omni-to-Omni Contrastive (OOC), the latter covering arbitrary cross-modal pairs without restriction.

A configurable data processor consolidates heterogeneous raw inputs (\textit{e.g.}, images, video frames, captions, ASR transcripts, OCR tokens) into a standardized set of modalities. For each training sample, we dynamically construct all valid query–target pairs according to predefined matching patterns that account for modality characteristics.  Modalities without valid patterns are excluded from the loss to mitigate noise propagation. Unlike static formulations that fix the number of losses per dataset, our \emph{dynamic multi-pattern} matching evaluates all feasible query–target pairs for each sample within the same forward pass. This not only maximizes the utilization of extracted embeddings but also improves optimization stability, as reflected by smoother convergence in training loss according to our experiments.

\begin{figure}[t]
    \centering
    \includegraphics[width=0.85\linewidth]{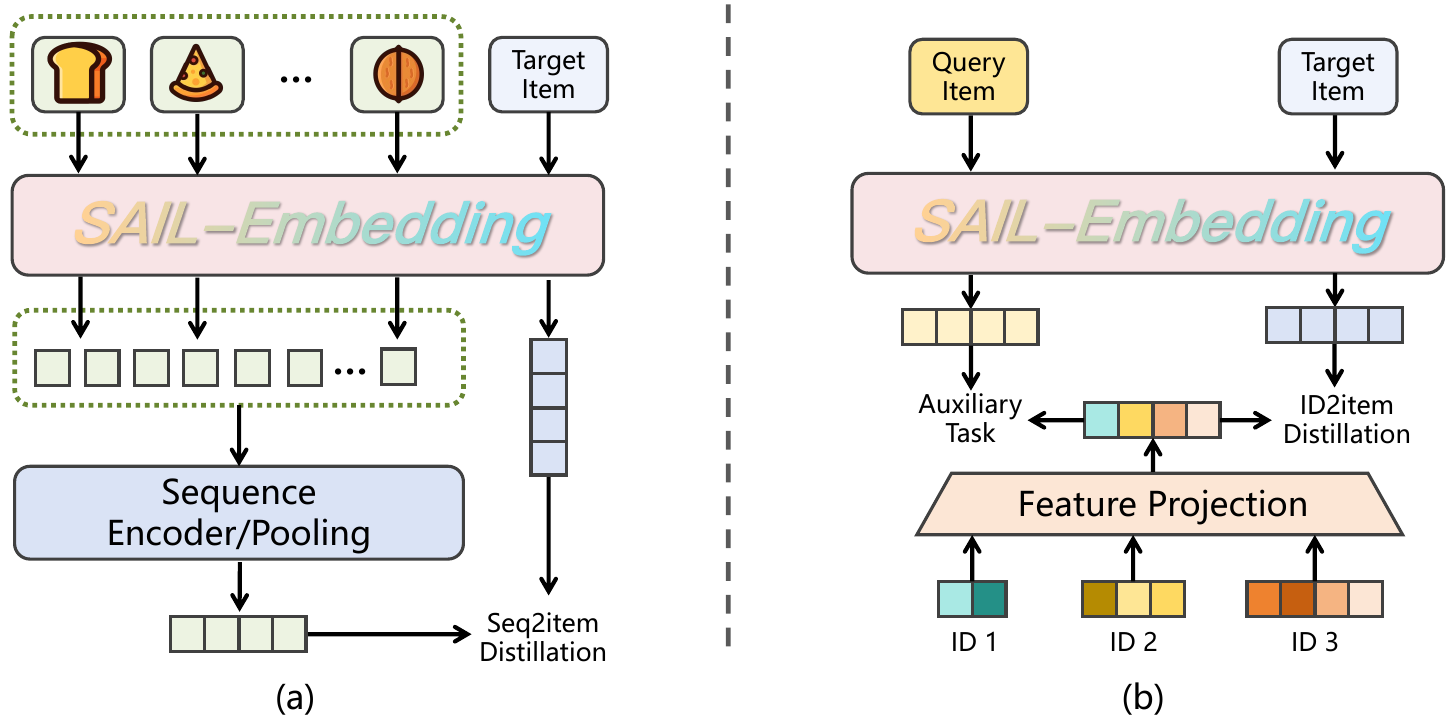}
    \caption{\textbf{Recommendation Enhancement Training}. We implement the (a) sequence-to-item distillation and (b) ID-to-item distillation to enhance the \modelname's collaboration-aware capabilities.
  }
    \label{fig:distill}
\end{figure}

\subsubsection{Collaboration-aware Recommendation Enhancement Training}
To address the demands of downstream recommendation scenarios for capturing user interests and to overcome the limitations of existing single-video content understanding, inspired by~\cite{deng2024end}, we propose a collaboration-aware recommendation enhancement training strategy.

\textbf{Sequence-to-Item Distillation.}
In data construction, we select user query sequence-to-target video pairs from four perspectives. Specifically, we first choose the users' historical video viewing sequence (1k) and filter it based on labels indicating positive interaction behaviours. The most recent item in the sequence is designated as the target video.

\begin{enumerate}
    \item \textbf{Content-aware Single-Peak Interest Modelling}: This part of the sequence data is retained by satisfying videos with at least three positive behaviours to mine items with compact interest distributions. Subsequently, items are further filtered based on similarity in content representations.
    \item \textbf{Content-aware Multi-Peak Interest Modelling}: This part of the sequence data is retained by satisfying videos with any positive behaviour to mine items with widely distributed interests. Then, we select sequence items and the target item based on the Jaccard coefficient, choosing those with behaviour label thresholds greater than 0.5.
    \item \textbf{Collaboration-aware Single-Peak Interest Modelling}: This part of the sequence data is retained by satisfying at least three positive behaviours. The Jaccard coefficient is still employed to measure behaviour consistency with the target item. We perform label-based clustering to select items that satisfy the original user interaction distribution.
   \item  \textbf{Collaboration-aware Multi-Peak Interest Modelling}: This part of the sequence data is retained by satisfying at least one positive behaviour. Subsequently, we select items corresponding to the proportional distribution of clustered videos as the final sequence data.
\end{enumerate}

As shown in Figure~\ref{fig:distill}(a), we perform contrastive learning by using the representations of the query sequence and the target video. In practice, we find two effective sequence modeling approaches: the mean pooling and the sequence encoder. The former inputs each video in the query sequence into an embedding model to obtain a sequence of query video embeddings, which are then aggregated via pooling. The latter employs a three-layer transformer module to construct a sequence encoder, using a special token to extract the sequence's overall representation.

\textbf{ID-to-Item Distillation.}
In Figure~\ref{fig:distill}(b), we attempt to directly align the model's output with the recommendation-side representations. In practice, we jointly utilize multiple ID embeddings of each item from the recommender system, aligning them with the omni-modal representation via feature projection to perform feature distillation. Simultaneously, we introduce an auxiliary i2i retrieval task and perform optimization in a multi-task manner to prevent the representation distribution from overly sacrificing content-aware perception capabilities.
The above joint training enhances the model's ability to aggregate video content and mine user interests, thereby better adapting to recommendation scenarios.

\section{Experiments and Results}

\subsection{Configuration Settings}
SAIL-Embedding is trained across multiple NPUs, using DeepSpeed ZeRO2~\cite{rajbhandari2020zeromemoryoptimizationstraining} and gradient checkpointing strategies.
In the modality components, the latent dimension, number of layers, and sampling depth of the visual perceiver are set to 1024, 16, and 6 respectively, with the self-attention enabled. Audio tokens extracted from the CLAP model~\cite{elizalde2023clap} have an original dimension of 512 and are mapped via an adapter to a 1536-dimensional alignment hidden state.
For the model optimization, we employ the FusedAdam optimizer, which supports mixed-precision computation and gradient fusion, significantly enhancing computational efficiency while maintaining training accuracy. The initial learning rate and weight decay coefficient are set to 1e-5 and 1e-4 to mitigate overfitting risks. The learning rate schedule employs a cosine annealing strategy, with the variation range constrained between 1e-5 and 6e-6 to prevent gradient explosion. In addition, a warm-up strategy is applied during the initial phase to stabilize optimization, and gradient clipping is used to further ensure training stability by limiting gradient norms.
In the seq2item distillation, we default to the mean pooling method and set the sequence length to 10 for efficient training. The sequence dataset representing diverse interests comprises 11M samples. The main task dataset in the ID2item distillation contains 220M samples, while the auxiliary task dataset holds 20M samples.

We compare the CLIP-based model~\cite{radford2021learning}  and the standard VLM-based approach~\cite{zhang2024gme} mainly on a wide range of item-to-item (i2i) tasks. Both are fine-tuned on the training data in Table~\ref{tab_data} to ensure fair comparisons. Additionally, performance comparisons on query-to-item (q2i)  tasks further involve unimodal embedding models, \textit{i.e.}, Doubao-Embedding and Qwen3-Embedding~\cite{zhang2025qwen3}.

\subsection{Evaluation Metrics}

In recommendation scenarios, traditional evaluation metrics for multimodal embeddings often fail to capture their true business impact. A multimodal embedding model may achieve strong standalone performance but contribute little to downstream user engagement. Moreover, recommendation systems evolve rapidly, where inefficient trial-and-error without robust pre-deployment evaluation risks suboptimal online performance. These challenges motivate the design of a multi-dimensional evaluation framework tailored to multimodal embedding models in recommendation tasks.

Given a dataset 
$\{q_i, t_i\}_{i=1}^N$, where $q_i$ denotes a query item and $t_i$ its matched target, a multimodal encoder $\mathcal{E}$ produces embeddings $\mathcal{E}(q_i), \mathcal{E}(t_i)$. The objective is to assess whether the resulting embedding space $(\mathcal{Q}, \mathcal{T})$ exhibits desirable properties: \textbf{structural similarity, stable ranking, and strong discriminability}. Our evaluation protocol consists of four complementary dimensions:

\begin{table}[t]
\caption{\textbf{Performance Comparison on i2i Tasks.} We consider four categories of realistic applications, including content understanding, search, and collaborative perception scenarios.}
\resizebox{0.8\textwidth}{!}{%
\begin{tabular}{cclcccc}
\toprule[1.5pt]
                                                                                      & \multicolumn{2}{c}{}                                                      &                                         & \multicolumn{3}{c}{Models}                          \\
\multirow{-2}{*}{Scenarios}                                                            & \multicolumn{2}{c}{\multirow{-2}{*}{Task}}                                & \multirow{-2}{*}{Metric}                & CLIP-based Model~\cite{radford2021learning} & VLM-based Model~\cite{zhang2024gme} & SAIL-Embedding \\ \midrule
                                                                                      & \multicolumn{2}{c}{}                                                      & {Recall@50}  & 17.83            & 19.99           & \textbf{20.76} \\
                                                                                      & \multicolumn{2}{c}{\multirow{-2}{*}{University Sub-i2i}}                 & {Recall@100} & 21.27            & \textbf{24.68}  & 24.46 \\
                                                                                      & \multicolumn{2}{c}{}                                                      & {Recall@50}  & 36.64            & 60.71           & \textbf{62.26} \\
                                                                                      & \multicolumn{2}{c}{\multirow{-2}{*}{Travel Sub-i2i}}                     & {Recall@100} & 45.87            & 70.61           & \textbf{72.40}  \\
                                                                                      & \multicolumn{2}{c}{}                                                      & {Recall@50}  & 75.98            & 81.32           & \textbf{89.08} \\
                                                                                      & \multicolumn{2}{c}{\multirow{-2}{*}{Film-i2i}}                           & {Recall@100} & 80.40            & 84.79           & \textbf{91.86} \\
                                                                                      & \multicolumn{2}{c}{}                                                      & {Recall@50}  & 95.31            & \textbf{98.75}  & 98.13          \\
                                                                                      & \multicolumn{2}{c}{\multirow{-2}{*}{Store Visit (Restaurant)-i2i}}       & {Recall@100} & 96.69            & \textbf{99.07}  & 98.69          \\
                                                                                      & \multicolumn{2}{c}{}                                                      & {Recall@50}  & 16.90            & 22.09           & \textbf{26.67} \\
                                                                                      & \multicolumn{2}{c}{\multirow{-2}{*}{Store Visit (City)-i2i}}             & {Recall@100} & 22.00            & 27.65           & \textbf{32.10}  \\
                                                                                      & \multicolumn{2}{c}{}                                                      & {Recall@50}  & 58.00            & 62.19           & \textbf{62.35} \\
                                                                                      & \multicolumn{2}{c}{\multirow{-2}{*}{Hot Topic-i2i}}                      & {Recall@100} & 74.43            & \textbf{80.56}  & 77.98          \\
                                                                                      & \multicolumn{2}{c}{}                                                      & {Recall@50}  & 71.38            & 70.17           & \textbf{78.07} \\
                                                                                      & \multicolumn{2}{c}{\multirow{-2}{*}{Comment-i2i}}                        & {Recall@100} & 82.16            & 80.57           & \textbf{87.62} \\
                                                                                      & \multicolumn{2}{c}{}                                                      & {Recall@50}  & 74.38            & \textbf{85.30}  & 84.55          \\
                                                                                      & \multicolumn{2}{c}{\multirow{-2}{*}{Music Play-i2i}}                     & {Recall@100} & 79.90            & \textbf{89.25}  & 88.72          \\
                                                                                      & \multicolumn{2}{c}{}                                                      & {Recall@50}  & 71.36            & \textbf{82.52}  & 82.08          \\
                                                                                      & \multicolumn{2}{c}{\multirow{-2}{*}{Music Gameplay-i2i}} & {Recall@100} & 77.25            & \textbf{86.98}  & 86.72          \\
                                                                                      & \multicolumn{2}{c}{}                                                      & {Recall@50}  & 67.78            & 71.15           & \textbf{72.83} \\
                                                                                      & \multicolumn{2}{c}{\multirow{-2}{*}{Game Tag-i2i}}                       & {Recall@100} & 72.15            & 74.92           & \textbf{76.93} \\
                                                                                      & \multicolumn{2}{c}{}                                                      & {Recall@50}  & 26.54            & 45.97           & \textbf{52.03} \\
                                                                                      & \multicolumn{2}{c}{\multirow{-2}{*}{Brand Vehicle-i2i}}                  & {Recall@100} & 33.15            & 52.22           & \textbf{57.34} \\
                                                                                      & \multicolumn{2}{c}{}                                                      & {Recall@50}  & 36.01            & 66.31           & \textbf{72.54} \\
                                                                                      & \multicolumn{2}{c}{\multirow{-2}{*}{Brand Phone-i2i}}                    & {Recall@100} & 46.84            & 76.68           & \textbf{81.64} \\
                                                                                      & \multicolumn{2}{c}{}                                                      & {Recall@50}  & 71.00            & 79.31           & \textbf{82.56} \\
                                                                                      & \multicolumn{2}{c}{\multirow{-2}{*}{Spot-i2i}}                           & {Recall@100} & 75.23            & 81.72           & \textbf{84.64} \\
                                                                                      & \multicolumn{2}{c}{}                                                      & {Recall@50}  & 67.36            & \textbf{76.12}  & 75.52          \\
                                                                                      & \multicolumn{2}{c}{\multirow{-2}{*}{Summary-i2i}}                       & {Recall@100} & 77.48            & \textbf{85.60}  & 85.56          \\
                                                                                      & \multicolumn{2}{c}{}                                                      & {Recall@50}  & 31.94            & 48.33           & \textbf{74.87} \\
                                                                                      & \multicolumn{2}{c}{\multirow{-2}{*}{Sub-i2i}}                            & {Recall@100} & 44.59            & 61.51           & \textbf{81.96} \\
                                                                                      & \multicolumn{2}{c}{}                                                      & {Recall@50}  & \textbf{57.27}  & 49.66           & 56.74          \\
\multirow{-38}{*}{\begin{tabular}[c]{@{}c@{}}Content \\ Understanding\end{tabular}}   & \multicolumn{2}{c}{\multirow{-2}{*}{ID Fusetag-i2i}}                     & {Recall@100} & \textbf{66.99}  & 60.13           & 67.45          \\ \midrule
                                                                                      & \multicolumn{2}{c}{}                                                      & {Recall@50}  & 56.72            & 77.45           & \textbf{80.50}  \\
                                                                                      & \multicolumn{2}{c}{\multirow{-2}{*}{Video Search-i2i}}                   & {Recall@100} & 62.23            & 80.90           & \textbf{84.01} \\
                                                                                      & \multicolumn{2}{c}{}                                                      & {Recall@50}  & 39.41            & 65.79           & \textbf{68.43} \\
\multirow{-4}{*}{Search}                                                              & \multicolumn{2}{c}{\multirow{-2}{*}{Search-i2i}}                         & {Recall@100} & 44.63            & 70.58           & \textbf{73.00}    \\ \midrule
                                                                                      & \multicolumn{2}{c}{}                                                      & {Recall@50}  & 49.76            & 42.52           & \textbf{52.46} \\
                                                                                      & \multicolumn{2}{c}{\multirow{-2}{*}{RSDF-i2i}}                            & {Recall@100} & 55.79            & 48.44           & \textbf{59.06} \\
                                                                                      & \multicolumn{2}{c}{}                                                      & {Recall@50}  & 53.47            & 66.06           & \textbf{69.17} \\
                                                                                      & \multicolumn{2}{c}{\multirow{-2}{*}{Copair-i2i}}                         & {Recall@100} & 70.6             & 72.70           & \textbf{75.57} \\
                                                                                      & \multicolumn{2}{c}{}                                                      & {Recall@50}  & 47.44            & 52.55           & \textbf{54.10}  \\
\multirow{-6}{*}{\begin{tabular}[c]{@{}c@{}}Collaborative\\  Perception\end{tabular}} & \multicolumn{2}{c}{\multirow{-2}{*}{Live-i2i}}                           & {Recall@100} & 56.69            & 62.60           & \textbf{63.80}  \\ \bottomrule[1.5pt]
\end{tabular}%
}
\label{tab_i2i}
\end{table}

\begin{table}[t]
\caption{\textbf{Performance Comparison on q2i Tasks.} We consider retrieval and classification applications across nine tasks.}
\resizebox{\textwidth}{!}{%
\begin{tabular}{clccccccc}
\toprule[1.5pt]
\multicolumn{2}{c}{Task}                                     & Metric                                             & Doubao-Embed.  & Qwen3-Embed.-4B       & Qwen3-Embed.-8B~\cite{zhang2025qwen3}       & CLIP-based Model~\cite{radford2021learning}        & VLM-based Model~\cite{zhang2024gme}         & SAIL-Embedding                   \\ \midrule
\multicolumn{2}{c}{\multirow{2}{*}{Short Video-q2i}}       & Recall@50              & 61.93                   & 71.73                   & 72.42                   & 74.16                   & 78.53                   & \textbf{86.53}                   \\
\multicolumn{2}{c}{}                                         & Recall@100            & 64.01                   & 75.17                   & 75.77                   & 76.21                   & 80.66                   & \textbf{88.54}                   \\ \midrule
\multicolumn{2}{c}{\multirow{1}{*}{Decision-q2i}}            & AUC & 74.51 & 69.79 & 69.39 & 65.18 & 67.34 & \textbf{82.44} \\
\multicolumn{2}{c}{\multirow{1}{*}{Longtail-q2i}}          & AUC & 85.84 & 83.90  & 83.96 & 83.18 & 84.02 & \textbf{91.22} \\
\multicolumn{2}{c}{\multirow{1}{*}{Search Longtail-q2i}}   & AUC & 76.86 & 74.90  & 75.34 & 74.82 & 74.99 & \textbf{83.31} \\
\multicolumn{2}{c}{\multirow{1}{*}{Unbiased-q2i}}          & AUC & 88.67 & 86.03 & 86.11 & 88.62 & 88.06 & \textbf{93.86} \\
\multicolumn{2}{c}{\multirow{1}{*}{Unbiased Longtail-q2i}} & AUC & 85.17 & 82.01 & 82.32 & 85.06 & 85.86 & \textbf{89.10}  \\
\multicolumn{2}{c}{\multirow{1}{*}{Biased-q2i}}            & AUC & 87.79 & 85.92 & 86.10  & 88.65 & 89.02 & \textbf{91.65} \\ \midrule
\multicolumn{2}{c}{\multirow{2}{*}{Live Summary2i}}      & Recall@50             & 66.31                   & 70.59                   & 69.79                   & 78.25                   & 81.55                   & \textbf{84.33}                   \\
\multicolumn{2}{c}{}                                         & Recall@100            & 69.31                   & 74.45                   & 73.85                   & 82.55                   & 85.62                   & \textbf{88.16}                   \\
\multicolumn{2}{c}{\multirow{2}{*}{Live-q2i}}              & Recall@50             & 59.91                   & 64.23                   & 64.22                   & 74.24                   & 73.91                   & \textbf{79.08}                   \\
\multicolumn{2}{c}{}                                         & Recall@100            & 64.76                   & 69.78                   & 69.57                   & 79.89                   & 79.34                   & \textbf{84.38}                   \\ \bottomrule[1.5pt]
\end{tabular}%
}
\label{tab_q2i}
\end{table}

\begin{enumerate}

    \item \textbf{Retrieval Recall.} We compute recall@k to quantify retrieval performance:
\[
\text{Recall@k} = \frac{\text{Number of relevant items retrieved in top-k}}{\text{Total number of relevant items}}.
\]
This is applied to i2i and q2i benchmarks, with $k \in \{1, 10, 25, 50, 100\}$.

    \item \textbf{Positive–Negative Separability.} For each query, we compute cosine similarity with its ground-truth target and with randomly sampled negatives. The distributional gap between positive and negative similarities reflects the model’s discriminative capability.

    \item \textbf{Group-wise Clustering Consistency.} 
We cluster the embeddings of positive and negative samples offline and adopt \emph{Normalized Mutual Information} (NMI) to measure consistency:  
\[
\text{NMI}(L_q,L_t) = \frac{2 I(L_q; L_t)}{H(L_q) + H(L_t)},
\]
where $L_q$ and $L_t$ denote cluster assignments, $I(\cdot;\cdot)$ is mutual information, and $H(\cdot)$ denotes entropy. 
A higher NMI indicates that positive and negative samples are distinguished by similar feature dimensions in the multimodal semantic space. 
For example, both positive and negative samples may cluster into the category ``anime,'' yet user preferences differ with respect to this content. 
This consistency highlights two key properties:  
(1) \emph{Effectiveness of modality information} — multimodal features capture the actual decision basis underlying user behavior (e.g., a user may like or dislike anime due to its ``art style'');  
(2) \emph{Strong interpretability of representations} — the observed consistency suggests that multimodal features bear a potential causal relationship with user behavior, rather than reflecting only superficial correlations.

    \item \textbf{Ranking Consistency.} We assess whether query–target similarity rankings remain stable across spaces. Top-$k$ overlap and Kendall’s $\tau$ are used:
\[
\tau = \frac{C - D}{C + D},
\]
where $C$ and $D$ are the numbers of concordant and discordant pairs, respectively. Higher $\tau$ indicates better preservation of relative semantic ordering.

    \item \textbf{Bijective Alignment Test.} We verify whether embeddings exhibit near one-to-one mapping. Specifically, for each query, we retrieve its most similar target, and from this target, retrieve back its most similar query. The proportion of cases where the indices match quantifies embedding fidelity and bijective consistency.

    \item \textbf{AUC Measurement.} Additionally, we report the AUC metric on recommendation-oriented tasks.

\end{enumerate}

Our comprehensive evaluation suite provides not only a quantitative measure of embedding quality but also actionable diagnostics on the suitability of multimodal embedding models for integration into large-scale recommendation systems.

\subsection{Evaluation Dataset Introduction}
We curate several evaluation benchmarks to assess our models. Below is the detailed introduction:

For item2item evaluation, we build three types of benchmarks:
\begin{itemize}
    \item \textbf{Content Understanding.} To evaluate the basic ability to capture semantic similarities, we integrate items with similar visual cues and textual information into pairs according to various rules. 
    Except for the test split of training datasets, we collect some out-of-domain i2i benchmarks to evaluate the generalizability of our model. 
    Leveraging the auto-tagging pipelines and expert models, we group items with the same brands or intellectual property, resulting in Brand Vehicle-i2i, Brand Phone-i2i, Film-i2i and Game Tag-i2i.
    Moreover, we employ MLLMs to summarize key information for items, and group them with the same n-grams or keywords, resulting in Keywords-i2i and Summary-i2i.
    
    \item \textbf{Search-Based.} 
    Taking user actions into consideration, we pair search-and-click items with the same queries. Specifically, Search-i2i consists of items with various forms, \textit{e.g.}, videos and photos. While video search-i2i only contains video items, which are mostly delivered on Douyin.
    
    \item \textbf{Collaborative Perception.} 
    To evaluate models' ability in recommendation scenarios, we collect item pairs based on the co-occurrence mechanism. We construct datasets for video items and live streaming items, respectively.
    Besides, we directly leverage dense features of the recommender system to compute the cosine similarity scores between items. Items with scores exceeding a particular score are reserved as positive pairs.

\end{itemize}

For query2item evaluation, we build two types of benchmarks:
\begin{itemize}
    \item \textbf{Query2item Retrieval Datasets.} Short Video-q2i is an in-domain dataset which is similar to the Search-q2i dataset for training. Besides, we collect two out-of-domain benchmarks from live streaming scenario. 
    Live-q2i utilizes user input search text as queries while Live-summary2i uses the summarized texts of each item.

    \item \textbf{Query2item Classification Tasks.} 
    To assess the model's ability to discriminate similar queries and items, we further collect six embedding-based classification datasets from different downstream scenarios.
    For evaluation, we acquire cosine similarity scores between all queries and items with their embeddings. The scores are regarded as the binary classification probabilities, which predict whether a query and an item are paired positives or unpaired negatives.
    The AUC metric is computed as the final result.
    In these datasets, all items contain only text features.
    
\end{itemize}

\subsection{Results Analysis}

\subsubsection{Item-to-Item Retrieval Task Comparison Results}

In Table~\ref{tab_i2i}, we present comparison results across 21 i2i tasks spanning four realistic application intents, including content understanding, search, and collaborative perception. We primarily perform omni-modal item retrieval. SAIL-Embedding significantly and consistently outperforms previous models in search and collaborative perception scenarios.
Compared to the CLIP-based model, SAIL-Embedding has richer open-world knowledge and feature fusion capabilities, demonstrating stronger multimodal understanding when handling complex and dynamic business scenarios. Compared to the VLM-based model, our model verifies the importance and effectiveness of the audio modality in multimedia retrieval driven by short videos. Furthermore, the well-designed training strategies also better eliminate the performance gaps of large-scale retrieval models in industrial applications.
Under the content understanding-oriented scenario, SAIL-Embedding demonstrates competitive performance across most tasks, suggesting the potential of the proposed embedding model in content-semantic learning.

\subsubsection{Query-to-Item Retrieval Task Comparison Results}
\begin{figure}[t]
    \centering
    \includegraphics[width=1\linewidth]{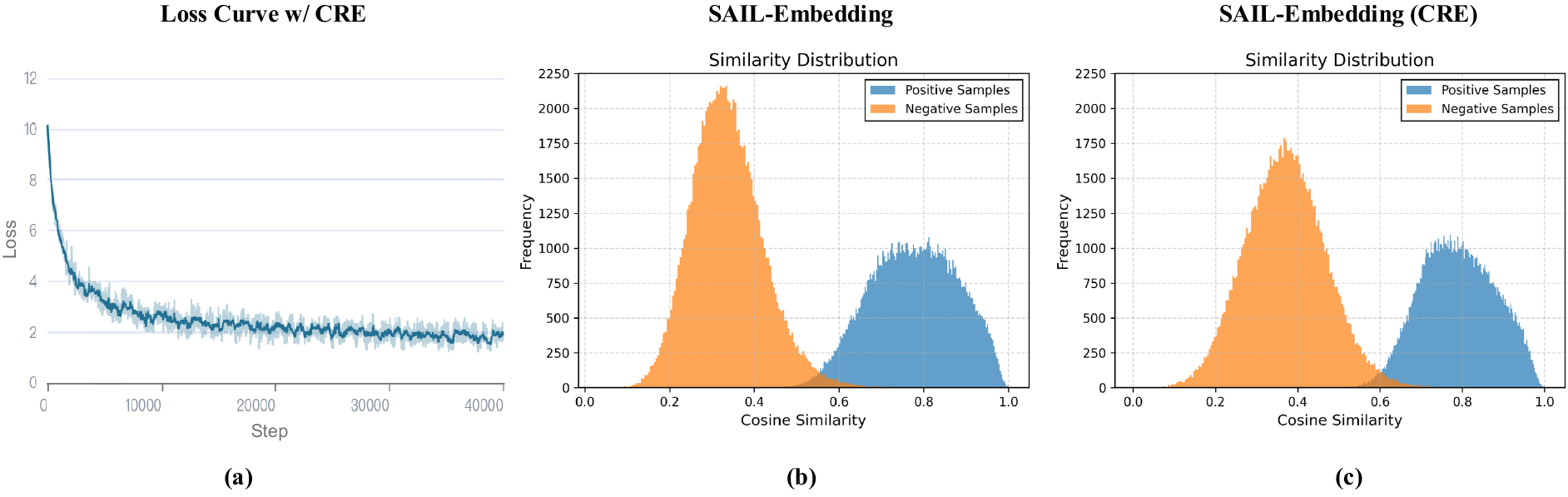}
    \caption{\textbf{Illustration of the Results with Collaboration-aware Recommendation Enhancement (CRE).} (a) Loss curve of the model fine-tuned with distilled ID embeddings. (b) Distribution of similarity scores for positive and negative samples in the original \modelname. (c) Distribution of similarity scores after applying CRE.
}
    \label{fig:cre}
\end{figure}

In Table~\ref{tab_q2i}, we further report results for various retrieval methods across 9 q2i tasks, encompassing both uni-modal and vision-language models. These tasks are categorized into retrieval-oriented and classification-oriented tasks, evaluated by Recall and AUC, respectively. SAIL-Embedding significantly outperforms previous methods across all metrics for every task. A plausible explanation is that the omni-modal architecture promotes unified semantic fusion and resolves modality ambiguity, leading to stronger cross-modal semantic understanding. Furthermore, the dataset-driven pattern matching and stochastic specialization training enhance SAIL-Embedding's cross-domain generalization capabilities and cross-task adaptability.

\subsubsection{Collaboration-aware Recommendation Enhancement Results}
Through Collaboration-aware Recommendation Enhancement (CRE), we adapt \modelname to better suit downstream recommendation tasks. 
To evaluate the effectiveness of this training phase, we report the results under different metrics in Figure~\ref{fig:cre} and Table~\ref{tab: cre}. 
As shown in Figure~\ref{fig:cre}, after incorporating Sequence/ID embedding distillation, the model converges stably after around 20k steps. 
We measure the separability illustrated in Figures~\ref{fig:cre}(b)\&(c). 
Before and after distillation, the model already exhibits good discrimination between positive and negative samples. 
However, with distillation, the overlap between positive and negative samples is further reduced and shifts rightward, verifying that the strategy effectively enhances the discriminative power of the base model, particularly for hard negatives with similarity scores around 0.5. 
In addition, the positive sample distribution becomes more compact, indicating that by jointly incorporating some ID embeddings, the model learns correlations among positive samples with higher confidence. 
This capability reflects exactly the kind of improvement expected in recommendation, where multimodal models should not only preserve separability but also strengthen relevance modeling for positive samples.


 \begin{table}[t]
\centering
\caption{\textbf{CRE Training Quantitative Results}. NMI, Ken., Inter., and Acc. represent Normalized Mutual Information in group-wise clustering consistency, Kendall’s coefficient, intersection proportion in ranking consistency, and the proportion of successful hits in the bijective alignment test, respectively. For Ken.*, the values are multiplied by 10K for clarity.} 
\resizebox{0.8\textwidth}{!}{%
\label{tab: cre}
\begin{tabular}{l|cccccccc}
\toprule[1.5pt]

\multicolumn{1}{c|}{Models}      & \multicolumn{1}{c}{NMI}                          & \multicolumn{1}{c}{Ken.*} & \multicolumn{1}{c}{Inter.} & Acc.  & \begin{tabular}[c]{@{}c@{}}Gid-i2i\\ R@50\end{tabular} & \begin{tabular}[c]{@{}c@{}}Gid-i2i\\ R@100\end{tabular} & \begin{tabular}[c]{@{}c@{}}Copair-i2i\\ R@50\end{tabular} & \begin{tabular}[c]{@{}c@{}}Copair-i2i\\ R@100\end{tabular} \\

\midrule
VLM-based Model       & 0.53                     & 68.6                                & 28.29                                     & 46.51 &     42.52                                                         &   48.44                                                            &      66.06                                                           &   72.70                                                             \\ 
VLM-based Model (CRE)                     & 0.61                   & 68.6                                & 29.62                                     & 47.22 &       46.34                                                      &   52.18                                                           &  64.78                                                              &   70.52                                                              \\
SAIL-Embedding        & 0.60 & 73.4                                & 29.73                                     & 48.28 & 52.46                                                       & 59.06                                                        & \textbf{69.17}                                                               &   \textbf{75.57}                                                             \\
SAIL-Embedding  (CRE) & \bf{0.65} & \bf{77.2}                                & \bf{31.63}                                     & \bf{48.94} & \bf{59.69}                                                       & \bf{66.37}                                                        & 66.83                                                               &  73.10    \\                                 \bottomrule[1.5pt]                         
\end{tabular}
    }
\end{table}

Quantitative results are summarized in Table~\ref{tab: cre}. 
Overall, we observe that the CRE strategy consistently improves performance across different metrics, both for VLM-based models and for our proposed \modelname. 
For the NMI, the ID-distilled version achieves a $+5\%$ gain, indicating that positive and negative samples become more separable in clustering, which facilitates learning of user preference boundaries.  
For the ranking consistency, the distilled model improves the Kendall correlation by $+3.8\%$ and the Intersection metric by $+1.9\%$, demonstrating enhanced consistency in retrieval orderings and suggesting stronger downstream recommendation performance.  
For the bijective alignment test, the distilled model improves by $+0.66\%$, showing that incorporating ID embeddings encourages \modelname to form more stable matching structures, which may provide valuable guidance for both recall and ranking stages.

Another interesting finding is that CRE training significantly improves model performance on the Gid-i2i benchmark. For instance, the CRE version of \modelname achieves 7.23\% and 7.31\% improvements over the original model on Recall@50 and Recall@100 metrics. The VLM-based model also achieves an average improvement of 3.78\%. This observation confirms that training via seq2item and ID2item distillation can further refine the model's omni-modal representations to perceive collaborative semantics.
This is reasonable since Gid-i2i is a paired testing benchmark constructed from \texttt{gid} embeddings’ relevance, which undergoes streaming updates during online execution.
Additionally, we observe some performance degradation on the Copair-i2i benchmark constructed based on item-based representation similarity. We consider this variation tolerable to balance the differing application requirements of content-oriented and collaborative behavior-oriented scenarios in industrial settings.

\subsubsection{Sequence Modeling Evaluation Results}
To verify the model's understanding of sequence semantics after sequence distillation, we build two evaluation sets from user historical viewing sequences to target videos, named \textit{Vanilla Seq2item} and \textit{Filtered Seq2item}. In the former, we collect 100 historical videos viewed by the user in chronological order. The most recent video is selected as the target, while the remaining 10 videos are randomly sampled from the sequence to form the user query sequence. 
In the \textit{Filtered Seq2item}, we extend the collected historical sequence to 1k videos and filter them based on content tags or clustered copair-ids to select items highly consistent with the target video as queries. The retained sequence length remains 10.

During implementation, we progressively perform seq2item and ID2item distillation on the content-aware training of SAIL-Embedding. Experimental results are presented in Table~\ref{tab:seq}. The model after seq2item distillation achieves significant performance gains on the \textit{Filtered Seq2item}.
For instance, the Recall@10 and Recall@25 metrics improved by 2.67\% and 4.66\%, respectively. This observation indicates that the model's sequence comprehension and content aggregation capabilities have been enhanced. Following further ID2item distillation, SAIL-Embedding consistently achieves higher performance, demonstrating that the combined distillation strategy strengthens the model's
collaboration-aware capabilities in recommendation scenarios.
On the \textit{Vanilla Seq2item}, the model ultimately achieves a 4.76\% metric improvement. This phenomenon suggests that recommendation-enhanced learning enhances the model's ability to extract diverse user interests.

\begin{table}[t]
\centering
\caption{\textbf{Sequence Modeling Evaluation Results}. Both test sets evaluate the recall accuracy for the top-1, top-10, and top-25 ranked items. The best results are emphasized in \textbf{bold}.} 
\resizebox{0.8\textwidth}{!}{
\label{tab:seq}
\begin{tabular}{c|cccccc}
\toprule[1.5pt]
\multirow{2}{*}{Models} & \multicolumn{3}{c}{Filtered Seq2item} & \multicolumn{3}{c}{Vanilla Seq2item} \\
\cmidrule(lr){2-7}
& R@1 & R@10 & R@25 & R@1 & R@10 & R@25 \\
\cmidrule(lr){1-7}
SAIL-Embedding & 8.41 &19.35 &25.46 &4.35 &10.46 &14.33 \\
SAIl-Embedding (Seq2item Distillation) & 8.28 &22.02 &30.12 &4.84 &11.78 &17.79 \\
SAIl-Embedding (Seq2item + ID2item Distillation) & \textbf{9.04} &\textbf{23.77} &\textbf{32.41} &\textbf{5.24} &\textbf{13.36} &\textbf{19.09} \\
\bottomrule[1.5pt]                         
\end{tabular}
    }
\end{table}

\subsubsection{Systematic Ablation Studies}

\begin{figure}[t]
    \centering
    \includegraphics[width=1\linewidth]{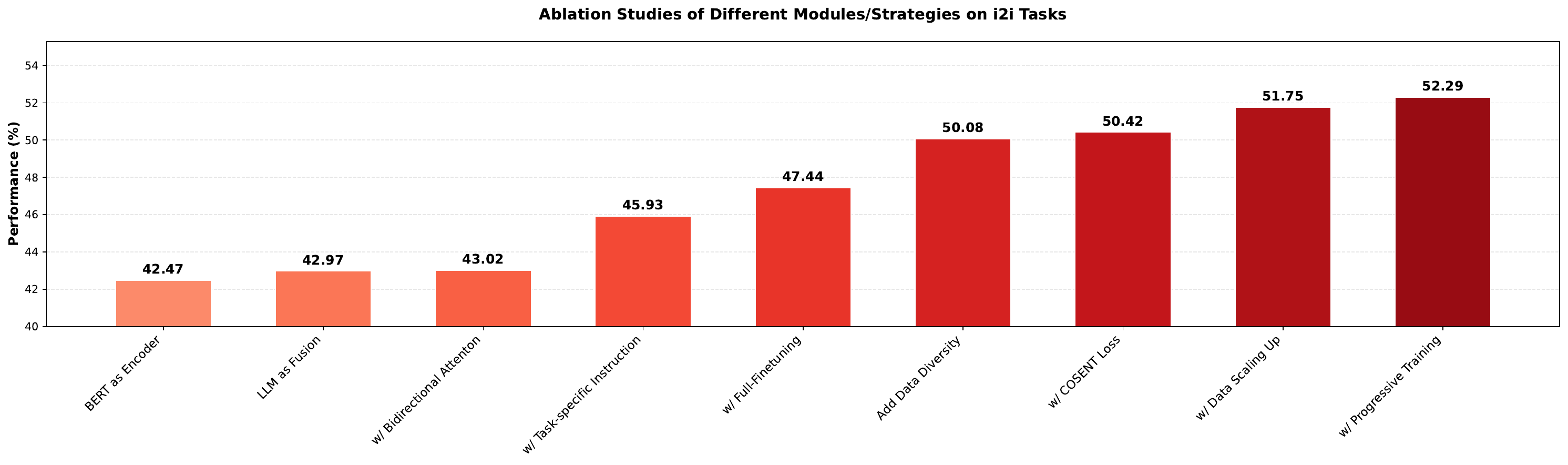}
    \caption{\textbf{Ablation Studies on i2i Tasks}. We systematically report on the effects of different modules and strategies on performance.
}
    \label{abl_i2i}
\end{figure}

\begin{figure}[t]
    \centering
    \includegraphics[width=1\linewidth]{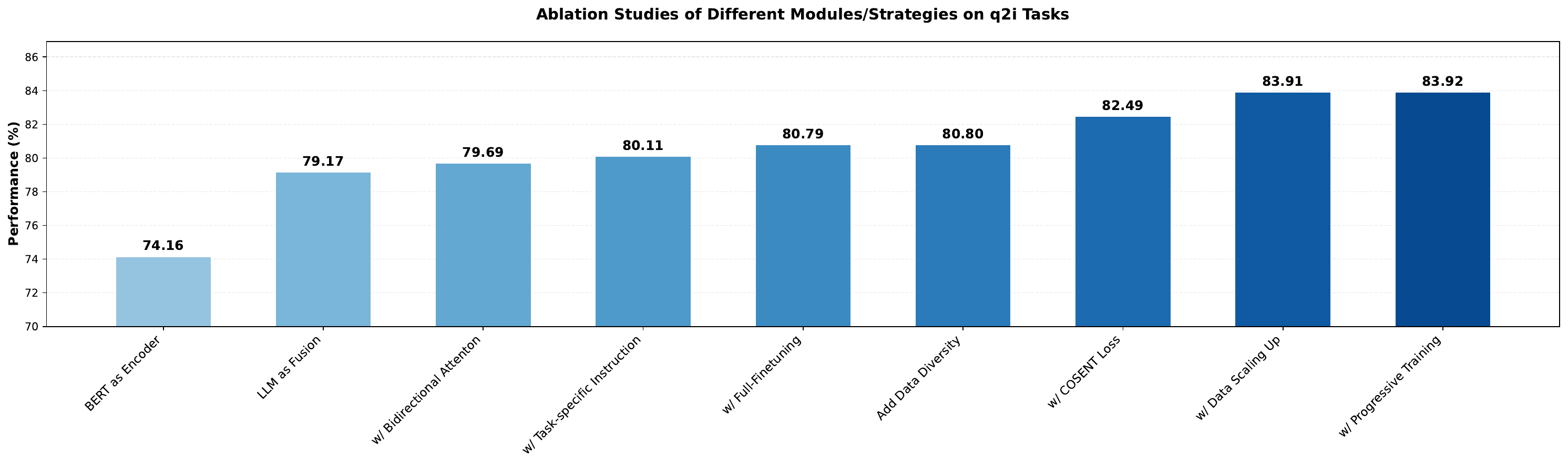}
    \caption{ \textbf{Ablation Studies on q2i Tasks}. We systematically report on the effects of different modules and strategies on performance.
}
    \label{abl_q2i}
\end{figure}

To comprehensively analyze the necessity of different components and strategies within the model, we conduct systematic ablation studies on subsets of i2i and q2i tasks. Figures~\ref{abl_i2i} and~\ref{abl_q2i} report average performance across different tasks for intuitive observation. We first establish a baseline performance where the BERT model~\cite{devlin2019bert} serves as an encoder for the text modality. Subsequently, the LLM is employed as a fuser to replace the traditional dual-tower pattern. 
We show consistent improvements across both tasks, with the q2i task achieving a significant gain of 5.01\%. This means the LLM enhances the model's ability to handle complex queries, boosting overall retrieval performance. When the original causal attention in the LLM is replaced with the bidirectional attention, positive incremental gains are demonstrated. The underlying reason is that the bidirectional attention captures global multimodal semantic dependencies, mitigating the semantic bias potentially introduced by the causal attention. This makes it more suitable for the representation embedding scenarios.

In the initial implementation, \modelname adopts a unified instruction format. Upon transitioning to task-specific instruction designs, we notice substantial gains. For instance, average performances on i2i and q2i scenarios increase by 2.62\% and 1.88\%, respectively. In practice, we observe that the instruction-based multi-task training paradigm guides the model to focus on distinct feature details. This enables adaptation to different task demands for feature expressiveness across various aspects of the omni-modal embeddings.
Additionally, we investigate the differences between using the Low-Rank Adaptation (LoRA)~\cite{hu2021lora} and full-parameter fine-tuning by fully unlocking parameters from the LLM. The results show that full parameter optimization effectively performs cross-modal semantic matching and retrieval between items, enhancing retrieval and classification performance in the q2i tasks.

On the one hand, when enhancing knowledge coverage and business data diversity during model training, \modelname achieves significant performance gains on i2i tasks while maintaining results across a wide range of q2i tasks. On the other hand, we discover that as data volume scales up, model performance increases accordingly, adhering to the effect of scaling laws. 
After incorporating the COSENT loss, the q2i tasks achieve a significant 1.69\% gain, implying that fine-grained ranking capabilities have been enhanced in text-oriented situations.
Furthermore, we modify the all-in-one training procedure used by the baseline to the progressive training strategy. Based on experimental results, we conclude that progressive training enables the model to acquire substantial domain knowledge in the early phases while establishing robust task dependencies and activating corresponding representational knowledge in subsequent phases.

\subsubsection{Extensive Online Experiments}

\begin{table}[t]
\centering
\caption{\textbf{Online Results in Recommender System}. Feed, Message Pushing, and Coldstart are different channels or modules of Douyin. Douyin-Selected is another application with its own recommendation pipeline.} 
\resizebox{0.65\textwidth}{!}{
\begin{tabular}{c|c|c|c}
\toprule[1.5pt]
Scenarios & Stage & Feature & Gain \\
\cmidrule(lr){1-4}
\multirow{4}{*}{Feed} & Recall & SID & LT30 + 0.01\% \\
 & Pre-Rank & SID &LT30 + 0.01\% \\
 & Rank & SID\&Embedding & Finish AUC + 0.1\% \\
 & Re-Rank & SID & LT30 + 0.01\% \\
 \cmidrule(lr){1-4}
 \multirow{2}{*}{Message Pushing} & Recall & SID \& Embedding & LT30 + 0.03\% \\
& Rank & SID & LT30 + 0.01\% \\
 \cmidrule(lr){1-4}
 Coldstart & Recall & SID \& Embedding & LT30 + 0.05\%\\
 \cmidrule(lr){1-4}
\multirow{2}{*}{Douyin-Selected} & Recall & SID & LT7 + 0.4\% \\
& Rank & Embedding & LT7 + 0.1\%\\
\bottomrule[1.5pt]                         
\end{tabular}%
\label{tab:rec}
}
\end{table}

We deploy our model in the real-world recommender system to further verify its effectiveness. 
Our model mainly provides two types of features for downstream use.:
\begin{enumerate}
    \item \textbf{Embeddings}. As mentioned before, our model compresses all information of an item, including video frames, title, OCR, ASR, along with the author's nickname and other textual tags into a dense embedding. This embedding can be used for similarity-based recall, or be used in modules like SIM.
    \item \textbf{Semantic ID (SID)}. Furthermore, we discretize the embeddings into semantic IDs since most recommenders prefer discrete features. We engage both clustering and vector quantization methods to acquire codebooks with different sizes. The discrete tokens can be used for decentralization as well as the features of candidate items.
\end{enumerate}

Experiments are conducted in various scenarios to verify the effectiveness of our method, and some typical results are shown in Table~\ref{tab:rec}.
Firstly, we find that the introduced features can benefit nearly all stages of the recommender system, including recall, pre-rank, rank and re-rank.
We believe that engaging features provided by the same embedding model can lead to better consistency through these sequential stages.
In addition, different scenarios such as coldstart and message pushing can both be enhanced, achieving \textbf{0.05\%} and \textbf{0.04\%} LT gain respectively. And our model can work on both Douyin and Douyin-Jingxuan, which are two different applications.
All these results have demonstrated that our model is both effective and generalizable for recommendation.

Moreover, we find that SIDs bring more gain than dense embeddings. The reason may come from two aspects: \textbf{(a)} SIDs are much easier to use in ruled-based methods than embeddings, such as rule-based decentralization. \textbf{(b)} SIDs can be encoded into trainable embeddings, like item ID, so that they can be further adapted for recommendation models. However, dense embeddings tend to bring more information than SIDs. To effectively leverage such information is worth discovering for future work.

\section{Conclusion}
In this work, we present SAIL-Embedding, an omni-modal embedding foundation model tailored for large-scale recommendation scenarios. By unifying vision, text, and audio modalities, SAIL-Embedding overcomes the limitations of limited modalities of existing multimodal methods. Our contributions include a dynamic hard negative mining strategy and an adaptive multi-source data balancing framework, which jointly enhance training robustness and representation quality. Furthermore, we design a content-aware progressive training procedure and a collaboration-aware recommendation enhancement module, enabling the model to capture both semantic content and collaborative behavioral signals. Extensive experiments across diverse item-to-item and query-to-item benchmarks demonstrate that SAIL-Embedding achieves state-of-the-art performance and superior generalization to real-world industrial applications. Beyond its empirical effectiveness, our systematic ablations validate the necessity of each proposed component. We believe SAIL-Embedding provides a scalable and versatile foundation for future multimodal retrieval and recommendation systems, and we envision extending this framework to broader downstream tasks such as video understanding, personalized content generation, and cross-domain knowledge transfer.

In future work, we plan to further enhance the integration of vision-language models (VLMs) into recommendation systems. 
First, we will explore training VLMs aligned with recommendation objectives and constructing generative tasks tailored for recommendation, enabling the model to acquire domain-specific knowledge at earlier stages and strengthen its recommendation capability. 
Second, during representation learning, we aim to better align model training with recommendation goals by mining more paired data from recommendation signals and behavioral feedback, thereby injecting user preferences into multimodal representations. 
Finally, we will investigate hard negative mining in recommendation scenarios to improve the robustness of representation learning.

\section{Contributor List}
All contributors are listed in alphabetical order by last name initial, with equal contributions
within each group; unless otherwise noted, all are members of the Douyin SAIL Team.

\textbf{Core Contributors:} \\
Lin Lin \quad Jiefeng Long \quad Zhihe Wan \quad Yuchi Wang~(CUHK MMLab) \quad Dingkang Yang \quad Shuang Yang \quad Yueyang Yao

\textbf{Contributors:} \\
Xu Chen \quad  Zirui Guo \quad  Shengqiang Li\quad   Weiran Li \quad  Hanyu Li \quad Yaling Mou \quad Yan Qiu \quad Haiyang Yu

\textbf{Project Leader:} \\
Xiao Liang

\textbf{Supervisors:} \\
Xiao Liang \quad Hongsheng Li~(CUHK MMLab) \quad Chao Feng

\section{Acknowledgments}
We sincerely thank Song Chen, Yaodi Du, Xinjie Huang, Shengyu Li, Zhenming Sun, Junyan Yao, Enwei Zhang, Tengfeng Zhang, Zeqin Zhang, and Sheng Zheng for their valuable support and contributions.

\clearpage

\bibliographystyle{plainnat}
\bibliography{main}



\end{document}